\documentclass[12pt,oneside, a4paper]{article}
\pdfoutput=1

\ifx\pdfoutput\undefined
\usepackage[dvips,bookmarks=false]{hyperref}	
\else
\usepackage{hyperref}	
\fi
\hypersetup{colorlinks,bookmarksopen,bookmarksnumbered,citecolor=blue,
linkcolor=black,pdfstartview=FitH,urlcolor=blue}

\usepackage{graphicx}
\usepackage{subfigure}
\usepackage{amssymb}
\usepackage{cite}
\usepackage{bm}
\usepackage{amsmath,amsthm}
\usepackage{cleveref}
\usepackage{mathrsfs}

\numberwithin{equation}{section}

\newcommand{\capdef}{}
\newcommand{\mycaption}[2][\capdef]{\renewcommand{\capdef}{#2}%
       \caption[#1]{{\footnotesize #2}}}

\newcommand{\iso}[2]{{\ensuremath{{}^{#2}}\ensuremath{\rm #1}}}
\newcommand{\parenbar}[1]{\overset{
            \raisebox{-0.15em}{\scalebox{.4}{\textbf{(}}}
            \raisebox{-0.3em}{{\hspace{.03em}--\hspace{.05em}}}
            \raisebox{-0.15em}{\scalebox{.4}{\textbf{)}}}} {#1}}

\begin{document}

\begin{titlepage}

\begin{center}

\vspace*{2cm}

{\Large\bf Sterile Neutrinos or Flux Uncertainties? --- \\[2mm]
  Status of the Reactor Anti-Neutrino Anomaly}
        \vspace{1cm}

\renewcommand{\thefootnote}{\fnsymbol{footnote}}

{\bf Mona Dentler}$^{a,}$\footnote[1]{modentle@uni-mainz.de}, 
{\bf \'Alvaro Hern\'andez-Cabezudo}$^{b,}$\footnote[2]{alvaro.cabezudo@kit.edu}, 
{\bf Joachim Kopp}$^{a,}$\footnote[3]{jkopp@uni-mainz.de},\\
{\bf Michele Maltoni}$^{c,}$\footnote[4]{michele.maltoni@csic.es},
{\bf Thomas Schwetz}$^{b,}$\footnote[5]{schwetz@kit.edu}
\vspace{5mm}

{\it $^a$PRISMA Cluster of Excellence and
             Mainz Institute for Theoretical Physics,\\
             Johannes Gutenberg-Universit\"{a}t Mainz, 55099 Mainz, Germany\\
$^b$Institut f\"ur Kernphysik, Karlsruher Institut f\"ur Technologie,
             76021 Karlsruhe, Germany\\
$^c$Instituto de F\'{\i}sica Te\'orica UAM/CSIC, Calle de
             Nicol\'as Cabrera 13--15,\\
             Universidad Aut\'onoma de Madrid, Cantoblanco, E-28049 Madrid, Spain             
}

\vspace{8mm} 

\abstract{
  The $\sim 3\sigma$ discrepancy between the predicted and observed
  reactor anti-neutrino flux, known as the reactor anti-neutrino
  anomaly, continues to intrigue.  The recent discovery of an
  unexpected bump in the reactor anti-neutrino spectrum, as well as
  indications that the flux deficit is different for different fission
  isotopes seems to disfavour the explanation of the anomaly in terms
  of sterile neutrino oscillations. We critically review this
  conclusion in view of all available data on electron (anti)neutrino
  disappearance.  We find that the sterile neutrino hypothesis cannot
  be rejected based on global data and is only mildly disfavored
  compared to an individual rescaling of neutrino fluxes from
  different fission isotopes. The main reason for this is the presence
  of spectral features in recent data from the NEOS and DANSS
  experiments. If state-of-the-art predictions for reactor fluxes are
  taken at face value, sterile neutrino oscillations allow a
  consistent description of global data with a significance close to
  $3\sigma$ relative to the no-oscillation case. Even if reactor
  fluxes and spectra are left free in the fit, a $2\sigma$ hint in
  favour of sterile neutrinos remains, with allowed parameter regions
  consistent with an explanation of the anomaly in terms of
  oscillations.}
\end{center}
\end{titlepage}

\renewcommand{\thefootnote}{\arabic{footnote}}
\setcounter{footnote}{0}

\setcounter{page}{2}
\tableofcontents

\section{Introduction}

Calculations of the anti-neutrino fluxes emmitted from nuclear
reactors performed in 2011 \cite{Mueller:2011nm, Huber:2011wv} have
led to an increased flux prediction compared to previous estimates
\cite{Schreckenbach:1985ep, Hahn:1989zr, VonFeilitzsch:1982jw,
  Vogel:1980bk}. This implies a deficit in the observed reactor
neutrino measurements compared to predictions, which is known
under the name ``reactor anti-neutrino anomaly''
(RAA)~\cite{Mention:2011rk}.  Using the published systematic errors on
the flux predictions, the significance of this anomaly is around $2.8
\sigma$. The anomaly can be explained by oscillations of electron
anti-neutrinos into a light sterile neutrinos with a mass-squared
difference of order 1~eV$^2$~\cite{Mention:2011rk}.  For recent
reviews on reactor neutrino flux calculations, and on possible caveats
with these calculations that could account for the anomaly, see
refs.~\cite{Hayes:2013wra, Hayes:2016qnu, Vogel:2016ted,
  Huber:2016fkt, Hayes:2017res}.  The neutrino oscillation hypothesis
is supported by an independent anomaly, namely a similar deficit of
neutrinos in experiments using an intense radioative source in
conjunction with gallium-based radiochemical detectors. This deficit
is usually referred to as the ``Gallium anomaly'' \cite{Acero:2007su,
  Giunti:2010zu}.

In this work we re-consider the sterile neutrino interpretation of the
reactor and gallium anomalies and update our analysis of this tension from
refs.~\cite{Kopp:2011qd, Kopp:2013vaa} (see also ref.~\cite{Gariazzo:2017fdh,
Collin:2016aqd}) in the light of the following recent experimental developments:
\begin{enumerate}
  \item 
  Precise measurements of the reactor anti-neutrino spectrum by modern
  experiments \cite{Seo:2016uom, Abe:2014bwa, An:2016srz} have
  revealed a spectral feature (``bump'') at neutrino energies around
  $E_\nu \sim $5~MeV, which is not predicted by the theoretical flux
  calculations. A compilation of results and a possible explanation in
  terms of detector energy scale non-linearity are presented in
  ref.~\cite{Mention:2017dyq}. The author of
  ref.~\cite{Huber:2016xis} concludes from the data that the likely
  source of this feature is the anti-neutrino flux from
  \iso{U}{235}\ fission. More discussions about possible origins of
  the bump can be found in ref.~\cite{Hayes:2015yka}.  While the
  origin of the bump is under debate and sheds some doubt on the
  reliability of flux calculations (or their error estimates), its
  presence cannot explain the RAA.

  \item
  Daya Bay~\cite{An:2016luf}, as well as the short-baseline reactor
  experiments NEOS \cite{Ko:2016owz} and
  DANSS \cite{Alekseev:2016llm, danss-moriond17} have presented new
  limits on sterile neutrino oscillations in the relevant parameter region.
  These new analyses rely on relative comparisons of measured spectra at different
  baselines and are therefore independent of flux predictions. While they find no
  clear evidence for oscillations, their observed spectra show some distortions
  which are consistent with the presence of a sterile neutrino in certain
  regions of parameter space. We will quantify the impact of these new
  results in relation to the previous RAA.

  \item
  Using the time evolution of the observed anti-neutrino rate and the
  knowledge of the isotopic composition of the reactor cores, the Daya
  Bay collaboration was able to determine the individual anti-neutrino
  fluxes from the four most important fissible isotopes \iso{U}{235},
  \iso{U}{238}, \iso{Pu}{239}, \iso{Pu}{241}~\cite{An:2017osx}. Their
  results suggest that the flux from \iso{U}{235}\ is the main source
  for the anomaly, while the one from \iso{Pu}{239}\ is consistent
  with the prediction. Fluxes from \iso{U}{238} and \iso{Pu}{241} are
  numerically less important.  Such a result would disfavour the
  sterile neutrino hypothesis, which predicts equal suppression of the
  fluxes from all isotopes. Below, we will quantify to what extent the
  Daya Bay measurement excludes a sterile neutrino explanation of the
  RAA in the context of global data.
\end{enumerate}

The hypothesis that the anomaly is due to a mis-prediction of the
\iso{U}{235} flux has already received support in the global analysis
from ref.~\cite{Giunti:2016elf}, predating the Daya Bay results of
ref.~\cite{An:2017osx}, and also in ref.~\cite{Giunti:2017nww}, which
includes the new Daya Bay data.  On the other hand, the authors of
ref.~\cite{Hayes:2017res} demonstrate that, when comparing the data to
a flux prediction based on nuclear data tables rather than measured
beta decay spectra, the anomaly is of similar magnitude for all
isotopes and therefore consistent with the sterile neutrino
hypothesis. In ref.~\cite{Giunti:2017yid} a combined analysis of the
new Daya Bay results~\cite{An:2017osx} with previous measurements of
the reactor neutrino rates has been performed, concluding that the
sterile neutrino hypothesis gives a fit of comparable quality to the
combined rate data as the \iso{U}{235}-only hypothesis. Below we will
present an analysis including previous rate measurements as well as
recent energy-spectral data, reaching a similar conclusion as the
authors of ref.~\cite{Giunti:2017yid}.

The outline of the paper is as follows: in \cref{sec:DBflux}, we
repeat the statistical analysis of the Daya Bay data from
ref.~\cite{An:2017osx}, comment on its interpretation, and carry out
additional statistical tests.  In \cref{sec:reactor}, we combine the
Daya Bay measurement of the individual isotopic fluxes with the other
globally available reactor data, paying special attention to the
impact of the new NEOS and DANSS results.  In \cref{sec:combined}, we
put our results in a wider context by also including $\nu_e$ 
disappearance data from gallium radiochemical experiments,
solar neutrinos, and accelerator experiments. We summarize and
conclude in \cref{sec:discussion}. Technical details of our
simulations are given in the appendix.

In this paper we restrict the analysis to the
$\protect\parenbar{\nu}_e$ disappearance sector, motivated by the
reactor and gallium anomalies. The implications of these results for
the LSND $\bar\nu_\mu\to\bar\nu_e$ signal \cite{Aguilar:2001ty} in the
context of global data on various oscillation channels will be
presented in a forth-coming publication.

\section{Daya Bay measurements of \iso{U}{235}\ and \iso{Pu}{239}\ fluxes}
\label{sec:DBflux}

In ref.~\cite{An:2017osx}, the Daya Bay collaboration has for the
first time presented independent measurements of the $\bar\nu_e$
fluxes from \iso{U}{235} and \iso{Pu}{239}\ fission. This analysis was
enabled by the precise knowledge of the isotopic composition of the
reactor fuel and its evolution with time, combined with the large
statistics of the Daya Bay near detectors.

The fuel composition is parameterized by the fractional contribution $F_{239}$
of \iso{Pu}{239} fissions to the total fission rate.
There is an approximate 1-to-1 correspondence between $F_{239}$ and the
fractional contributions of the other isotopes,
$F_{235}$ for \iso{U}{235}, $F_{238}$ for \iso{U}{238}, and $F_{241}$
for \iso{Pu}{241}, see figure~2 of \cite{An:2017osx}.  8 bins in $F_{239}$ are used.
Data are reported as effective inverse beta decay (IBD) yields
$\sigma$, given in units of cm$^2$ per fission.
We write the predicted IBD rates in each $F_{239}$ bin as
\begin{align}
  \sigma_\text{pred}^a = \sum_i P^i_\text{osc} \xi_i F_i^a \sigma_i^\text{HM} \,.
  \label{eq:sigma-pred}
\end{align}
Here, the index $i$ runs over the four fissible isotopes; $\sigma_i^\text{HM}$
is the IBD rate according to the Huber \& Mueller flux predictions
\cite{Mueller:2011nm, Huber:2011wv}; $F_i^a$ gives the effective contribution
of isotope $i$ to the total fission rate in the $a$-th $F_{239}$ bin ($a = 1
\dots 8$); $\xi_i$ are four pull parameters which allow each flux to deviate
from the predictions; and $P^i_\text{osc}$ is the averaged oscillation
probability at the Daya Bay Experimental Halls 1 and 2 (EH1 and EH2).  (Data
from EH3 is not used in this analysis.)
The predictions from ref.~\cite{Huber:2011wv} are
used for the isotopes \iso{U}{235}, \iso{Pu}{239}, \iso{Pu}{241}, and those
from ref.~\cite{Mueller:2011nm} are used for \iso{U}{238}.
$P^i_\text{osc}$ depends on the
oscillation parameters and has a small dependence on the isotope $i$ due to the
slightly different neutrino spectra for each isotope. In the region $\Delta
m^2_{41} \gtrsim 0.05$~eV$^2$, oscillations are averaged out completely,
$P^i_\text{osc}$ becomes independent of the isotope $i$ and acts just as a
global normalization factor, $P^i_\text{osc} \approx 1 - \frac{1}{2}
\sin^22\theta_{14}$. For smaller values of $\Delta m^2_{41}$ we take into
account the correct oscillatory behaviour.

As a test statistic for the analysis we use the $\chi^2$ function
\begin{align}
  \chi^2 = \sum_{a,b=1}^8 (\sigma_\text{obs}^a - \sigma_\text{pred}^a)
              V^{-1}_{ab} (\sigma_\text{obs}^b - \sigma_\text{pred}^b)
         + \chi^2_\text{flux}(\xi_i) \,.
  \label{eq:chisq-DBflux}
\end{align}
Here, $\sigma_\text{obs}^a$ and $\sigma_\text{pred}^a$ are the
observed and predicted IBD yields in the $a$-th $F_{239}$ bin.
The covariance matrix $V_{ab}$ includes statistical and
correlated systematic errors.  The covariance matrix as well as
$\sigma_\text{obs}^a$ and $F_i^a$ are taken from the
supplementary material of ref.~\cite{An:2017osx}.

The term $\chi^2_\text{flux}(\xi_i)$ constrains the nuisance parameters
$\xi_i$, and depending on the analysis we adopt different assumptions
for it. When we impose the Huber \& Mueller flux predictions (``fixed
fluxes''), this term takes into account the systematic uncertainties
on the fluxes as published in
refs.~\cite{Mueller:2011nm,Huber:2011wv}. We will also perform a
``free fluxes'' analysis, where $\xi_{235}$ and $\xi_{239}$ are allowed
to vary freely. In this analysis, $\chi^2_\text{flux}(\xi_i)$ still imposes
a weak $1\sigma$ constraint of 10\% relative to the Huber \& Mueller predictions
on the subleading isotopes ($\xi_{238}$ and $\xi_{241}$) to
avoid unphysical results.

\begin{table}
  \centering
  \begin{tabular}{lcccc}
    \hline\hline
    Analysis                 & $\chi^2_\text{min}/$dof & gof  & $\sin^2 2\theta_{14}^\text{bfp}$
                             & $\Delta\chi^2(\text{no osc})$ \\
    \hline
    fixed fluxes + $\nu_s$   & $9.8/(8-1)$ & 18\% & 0.11 & 3.9\\
    free fluxes (no $\nu_s$) & $3.6/(8-2)$ & 73\% &      &    \\
    \hline\hline
  \end{tabular}
  \mycaption{Fits to the Daya Bay flux measurements.  The ``fixed
    fluxes'' analysis assumes the Huber \& Mueller flux
    predictions~\cite{Mueller:2011nm,Huber:2011wv} (accounting for
    their quoted uncertainties) and includes $\bar\nu_e$ disappearance
    into sterile neutrinos $\nu_s$. We assume $\Delta m^2_{41} \gtrsim
    0.05$~eV$^2$, so that oscillations are in the averaging
    regime. For the ``free fluxes'' analysis, fluxes are allowed to
    vary freely, but $\theta_{14}$ is set to zero. The goodness-of-fit
    (gof) $p$-values are calculated by Monte Carlo simulation and agree roughly
    with the $\chi^2$ approximation.}
  \label{tab:DBflux}
\end{table}

In \cref{tab:DBflux} we show the results of our fit to the Daya Bay flux
data under the ``fixed fluxes'' and ``free fluxes'' assumptions. We assume
$\Delta m^2_{41} \gtrsim 0.05$~eV$^2$ so that the predictions
become independent of $\Delta m^2_{41}$ and the only relevant oscillation
parameter is $\theta_{14}$. For an analysis including sterile neutrinos,
the number of degrees of freedom is thus $8 - 1 = 7$. For the
``free fluxes'' analysis assuming no sterile neutrino ($\theta_{14} = 0$),
the number of degrees of freedom is 6, accounting for the two unconstrained
pull parameters $\xi_{235}$ and $\xi_{239}$.  We have checked by
explicit Monte Carlo simulation, that $\chi^2_\text{min}$ follows
indeed a $\chi^2$-distribution with 7 and 6 dof, respectively, to very
good accuracy. As is clear from \cref{tab:DBflux}, the hypothesis of free fluxes gives
a better fit to the data, with a goodness-of-fit (gof) $p$-value of
73\%. However, the sterile neutrino hypothesis also has an acceptable
gof with a $p$-value of 18\% and therefore cannot be rejected at
reasonable confidence. The best fit point has
$\sin^22\theta_{14} = 0.11$, and for fixed fluxes the no oscillation hypothesis is
disfavoured with $\Delta\chi^2 = 3.9$, corresponding to about
$2\sigma$ (1 dof) exclusion.

To quantify Daya Bay's preference for free fluxes compared to oscillations
into sterile neutrinos, we construct a test statistic
\begin{align}\label{eq:T}
  T = \chi^2_\text{min}(H_0) - \chi^2_\text{min}(H_1) \,,
\end{align}
which compares the two hypotheses. Here, we call $H_0$ the hypothesis
that the Huber \& Mueller fluxes are correct, but
$\bar\nu_e$ can disappear due to oscillations at the eV$^2$ scale; $H_1$
is the hypothesis that the predicted normalization of the four fluxes is not
trustworthy and should be left free. Hence, $H_0$ corresponds to the
``fixed fluxes'' analysis including oscillations, and $H_1$ to the
``free fluxes'' analysis without oscillations.
Note that to a good approximation $H_0$ is a subset of $H_1$, since
oscillations basically act as a global normalization. Under this
assumption we expect that $T$ follows a $\chi^2$ distribution with
1~dof. We have verified by Monte Carlo simulation that this is indeed
approximately true and holds independently of the true value of $\theta_{14}$, as
long as $\sin^22\theta_{14} \lesssim 0.6$. We find
\begin{align}\label{eq:Tobs}
  T_\text{obs} = 6.3 \,,\qquad p\text{-value} = 0.7\% ~ (2.7 \sigma) \,,
\end{align}
where the $p$-value is evaluated by Monte Carlo simulation (we obtain
1.2\% (2.5$\sigma$) in the approximation of a $\chi^2$ distribution with 1 dof).
Hence, the sterile neutrino hypothesis is disfavoured compared to the
``free fluxes'' hypothesis at the 99.3\% confidence level. This is in
qualitative agreement with the results of~\cite{An:2017osx}. The
reason for the slightly lower value for $T_\text{obs}$ in
\cref{eq:Tobs} compared to the value of 7.9 obtained in~\cite{An:2017osx}
is that our ``fixed fluxes'' analysis includes the uncertainties in
the Huber \& Mueller flux prediction for the four isotopes (encoded in
$\xi_i$ factors in \cref{eq:sigma-pred}), whereas
ref.~\cite{An:2017osx} does not.

In summary, while the Daya Bay flux measurements favour the ``free fluxes''
hypothesis over the sterile neutrino hypothesis, the latter still
provides a good fit to the data (gof of 18\%). Assuming the Huber \& Mueller
flux predictions to be correct, $\bar\nu_e$ disappearance is favored over the
no oscillation hypothesis at $2\sigma$. Therefore, we proceed with the sterile
neutrino analysis and combine the Daya Bay flux data with
all other reactor data.

\section{Combined Analysis of Reactor Neutrino Data}
\label{sec:reactor}

\subsection{Data Sets Used and Analysis Procedure}

Our analysis of reactor neutrino data is based on
ref.~\cite{Kopp:2013vaa}, where technical details and further
references can be found. Here we describe the main differences and
updates with respect to that analysis.
\Cref{tab:reactor} summarizes the data sets included in our global fit.
We distinguish experiments comparing the predicted and measured total neutrino
fluxes and experiments that use spectral information in the analysis.

Compared to ref.~\cite{Kopp:2013vaa}, we have added a 4th Krasnoyarsk
data point \cite{Kozlov:1999cs}, see ref.~\cite{Giunti:2016elf} for
details.  For RENO and DoubleChooz, we include in our analysis the
total rate measurements at the near detectors given in refs.~\cite{reno-EPS17} and 
\cite{Giunti:2016elf}, respectively. For RENO,
we also include the ratio of total rates at the near and far sites
from ref.~\cite{reno-Neutrino14}.  We do not include the RENO and
Double Chooz measurements of the neutrino spectrum, as a reliable
interpretation of these measurements in the context of sterile
neutrino models turned out to be difficult, and as their statistical
power is far inferior to that of Daya Bay.
The Daya Bay measurement of the isotope-dependent
neutrino fluxes \cite{An:2017osx} discussed in \cref{sec:DBflux} are included
as constraints on the flux normalizations in a consistent way, correlated
between all reactor data. In other words, the nuisance parameters $\xi_i$ in
\cref{eq:sigma-pred} are the same for each experiment, and the pull term
$\chi^2_\text{flux}(\xi_i)$ in \cref{eq:chisq-DBflux} is added only once to the
gloabl $\chi^2$.  Oscillation effects in Daya Bay that affect the extraction of
the fluxes are of course taken into account.

\begin{table}[t]
  \centering\small
  \begin{tabular}{lcclc} 
    \hline\hline 
    Experiment   & Ref. & \# Data & Comments & New? \\
    \hline
     Bugey-4      & \cite{Declais:1994ma}                 & 1 & rate & -- \\
     ILL          & \cite{Kwon:1981ua}                    & 1 & rate & -- \\
     G\"osgen     & \cite{Zacek:1986cu}                   & 3 & rates& -- \\
     Krasnoyarsk  & \cite{Vidyakin:1987ue,Vidyakin:1994ut,Kozlov:1999cs} & 4 & rates & -- \\
     Rovno88      & \cite{Afonin:1988gx}                  & 4 & rates & -- \\
     Rovno91      & \cite{Kuvshinnikov:1990ry}            & 1 & rate  & -- \\
     SRP          & \cite{Greenwood:1996pb}               & 2 & rates & -- \\
     RENO         & \cite{reno-EPS17,reno-Neutrino14} & 2 & rate at near detector + near-far rate ratio & -- \\
     Double Chooz & \cite{Giunti:2016elf}                 & 1 & rate at near detector & -- \\
     Daya Bay flux& \cite{An:2017osx}                     & 8 & individual fluxes for each isotope (EH1, EH2) & \checkmark \\
     \hline
     Bugey-3      & \cite{Declais:1994su}                 & 35 & spectra at 3 dist.\ with free bin-by-bin norm. & -- \\    
     NEOS         & \cite{Ko:2016owz,An:2016srz}          & 60 & spectral ratio of NEOS and DayaBay & \checkmark \\
     DANSS        & \cite{danss-moriond17}                & 30 & spectral ratio at two distances & \checkmark \\
     Daya Bay spect. & \cite{An:2016ses}                  & 70 & spectral ratios EH3/EH1 and EH2/EH1 & \checkmark \\
     \hline
     KamLAND      & \cite{Gando:2010aa}    & 17 & spectrum at very long distance & -- \\
    \hline\hline
  \end{tabular}
  \mycaption{Data from reactor neutrino experiments used in our
    analysis. Data are separated into integrated rate measurements,
    data on the neutrino energy spectrum, and the very-long baseline
    experiment KamLAND. The column ``\# Data'' gives the number of
    data points entering the corresponding $\chi^2$ function. The
    total number of data points is 239. The acronym ``EH'' stands for
    ``experimental hall'' in Daya Bay, with EH1, EH2 being the two
    near detectors halls and EH3 the far detector hall. The last
    column highlights the most recent data sets (since summer 2016).
    In the text, we refer to these data sets as ``new'', to the
    previous ones as ``old''.}
  \label{tab:reactor}
\end{table}

New data on the reactor anti-neutrino spectrum are included from the
Daya Bay, NEOS, and DANSS experiments. In ref.~\cite{An:2016luf} the
Daya Bay collaboration has presented constraints on sterile neutrino
mixing by fitting two ratios built out of the spectra recorded at the
three experimental halls.  We follow this strategy but use the larger
data sample from ref.~\cite{An:2016ses}. We fit the ratio of the
binned spectra at EH3/EH1 and EH2/EH1. Details of the analysis are
given in \cref{app:DBsterile}. The NEOS
collaboration~\cite{Ko:2016owz} has reported a high statistics
measurement of the anti-neutrino spectrum at a distance of 24~m from
the core of a 2.8~GW nuclear reactor. We include their results using
the ratio of the measured spectrum to the shape predicted from the
flux measured at the Daya Bay EH1 and EH2
detectors~\cite{An:2016srz}. Details of the analysis are given in
\cref{app:neos}. The DANSS collaboration has reported preliminary
results on the anti-neutrino event spectrum at distances of 10.7~m and
12.7~m from a reactor core~\cite{danss-moriond17}.  We include these
measurements by fitting the bin-by-bin ratio of the two spectra, see
\cref{app:danss} for details.  In all cases we have verified that we
can reproduce to good accuracy the results of the respective
experimental collaborations, when data are analysed under the same
assumptions.

\bigskip

As in \cref{sec:DBflux}, we will in the following present two different global
fits: one with fixed fluxes, in which we take the predicted anti-neutrino
fluxes and their uncertainties at face value; and one with free fluxes, in
which the flux from each fissible isotope is allowed to float independently. In
the case of fixed fluxes, the predictions from ref.~\cite{Huber:2011wv} are
used for the isotopes \iso{U}{235}, \iso{Pu}{239}, \iso{Pu}{241}, and those
from ref.~\cite{Mueller:2011nm} are used for \iso{U}{238}.  In this analysis we
always take into account the quoted systematic uncertainties on the
fluxes~\cite{Huber:2011wv}, including correlations between isotopes and energy
bins. These uncertainties are of order few~\%. In the fit with free fluxes, the
normalizations of the \iso{U}{235}\ and \iso{Pu}{239}\ fluxes are left
completely unconstrained, whereas for the subleading fluxes from \iso{U}{238}\
and \iso{Pu}{241}\ we impose a weak constraint of $\pm 20\%$ ($1\sigma$) in
order to avoid unphysical values. (Note that this is more conservative than the
$\pm 10\%$ uncertainty we used in \cref{sec:DBflux} to match Daya Bay's analysis.)
Thanks to the Day Bay flux measurement
\cite{An:2017osx} as well as the slightly different isotopic compositions of
the different reactor cores at which experiments have been conducted, the data
itself provides sufficient information on the flux normalizations, see e.g.,
refs.~\cite{Giunti:2016elf,Giunti:2017nww,Giunti:2017yid}. 

The predicted reactor neutrino \emph{spectra} are used neither in the
``fixed fluxes'' nor in the ``free fluxes'' fit. Instead, when using
spectral information, we always compare measured spectra at different
baselines.  Ignoring the predicted shape of the neutrino spectrum is
motivated by the unexplained bump at $E_\nu \sim 5$\ MeV, which
indicates that our theoretical understanding of the spectra is
incomplete\footnote{Predicted spectra are used to perform the energy integral for total rate measurements and for averaging each energy bin over the resolution function.}.  In Daya Bay, the spectral comparison is between the
different experimental halls; in DANSS, it is between measurements at
two different locations using the same (movable) detector; for NEOS,
the observed spectrum is compared to the measured spectrum from Daya
Bay; for Bugey-3, we have modifed our previous analysis
\cite{Grimus:2001mn, Kopp:2013vaa} by introducing a free pull
parameter for each energy bin and correlating it between the three
detector distances of 15, 40, and 95~m (this leads to $60-25=35$
degrees of freedom for Bugey-3).\footnote{Bugey-3 results are reported
  in 25 bins at 15 and 40~m and 10 bins at 95~m. We introduce 25 pull
  parameters, corresponding to the binning at 15 and 40~m. Then we take
  into account the fractional effect of each pull for each of the
  larger bins of the spectrum at 95~m.}  Let us remark that our
analysis of spectral data is somewhat over-conservative because we
allow the spectra to be deformed in an uncorrelated way between
different experiments, i.e., we do not correlate the energy bins
between different experiments (except for the NEOS--Daya Bay
comparison).

\bigskip
Concerning the oscillation physics, we use the full 4-flavour
disappearance probability. For our parameterization conventions see
ref.~\cite{Kopp:2013vaa}.  The parameters $\Delta m^2_{21},
\theta_{12}, \Delta m^2_{31}$ are fixed to the 3-flavour best fit
points, while $\theta_{13}$ is left free (since the used data
determine it with good accuracy).

\subsection{Results for the Combined Analysis of Reactor Data}

\begin{figure}[t]
  \centering
  \includegraphics[width=.67\textwidth]{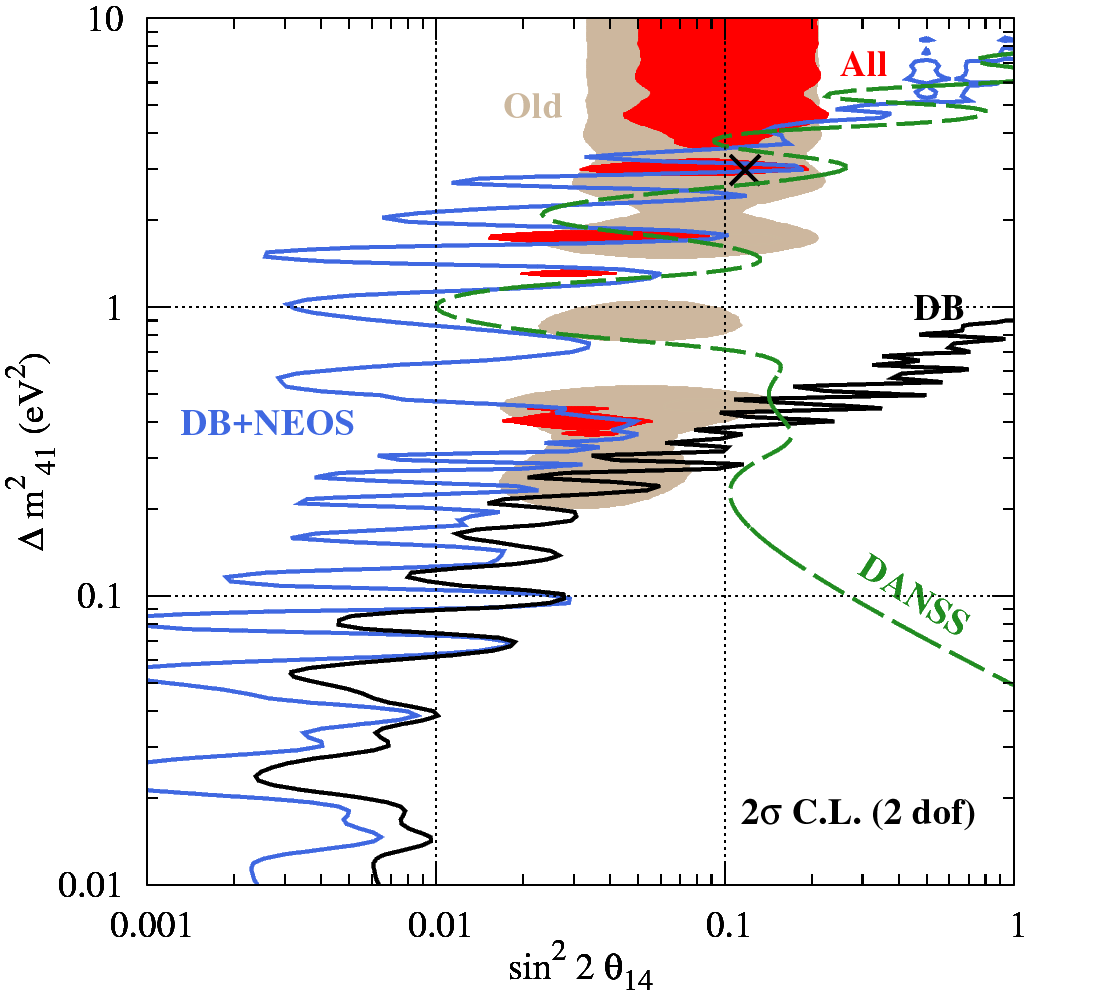}
  \mycaption{Allowed parameter regions at $2\sigma$ (2~dof) for the
    ``flux-fixed'' analysis, for the ``old'' data sample defined in
    \cref{tab:reactor} (khaki regions), for the
    DANSS~\cite{danss-moriond17}, Daya Bay spect.~\cite{An:2016ses},
    the combined Day Bay spect.\ + NEOS~\cite{Ko:2016owz} oscillation
    analyses, and the combined region of all data including also Daya
    Bay flux~\cite{An:2017osx} (red regions). The cross marks the best
    fit of the combined region.}
  \label{fig:old-vs-new}
\end{figure}

In \cref{fig:old-vs-new} we illustrate the impact of the recent
oscillation analyses from NEOS, DANSS, and the Daya Bay spectrum. In
khaki, we show the $2\sigma$ allowed parameter region in the $\sin^2
2\theta_{14}$ vs.\ $\Delta m_{41}^2$ plane, based on data predating
the summer conferences 2016. The corresponding data sets are marked
with ``--'' in the last column of \cref{tab:reactor}. The black and
green dashed contours show the new exclusion limits from Daya Bay and
DANSS, and the blue contours depict the limit from the combined NEOS
and Daya Bay spectral analysis.  Due to the relatively long baseline of the Daya
Bay detectors, these data constrain the region of $\Delta m^2_{41}
\lesssim 0.3$~eV$^2$, while both NEOS and DANSS are most sensitive in
the RAA region around few~eV$^2$.

As mentioned above, the NEOS analysis is based on the ratio of the
spectra in the NEOS detector to the one extracted from Daya Bay EH1
and EH2 data. When taking into account the $\Delta m^2_{41}$
dependence of the oscillations in the Daya Bay near detectors, NEOS
data lead to a closed regions with a best fit point below $\Delta m^2_{41}
\simeq 0.1$~eV$^2$, which is, however, excluded by the Daya Bay
spectral data at the far detector (EH3). Therefore, we decided to show
only the combined NEOS+Daya Bay (spectrum) constraint, in order to
avoid the effect of the minimum in the excluded region. The
complementarity of the two data sets is clearly visible, by comparing
the blue and black curves.

\begin{figure}[t]
  \centering
  \includegraphics[width=.45\textwidth]{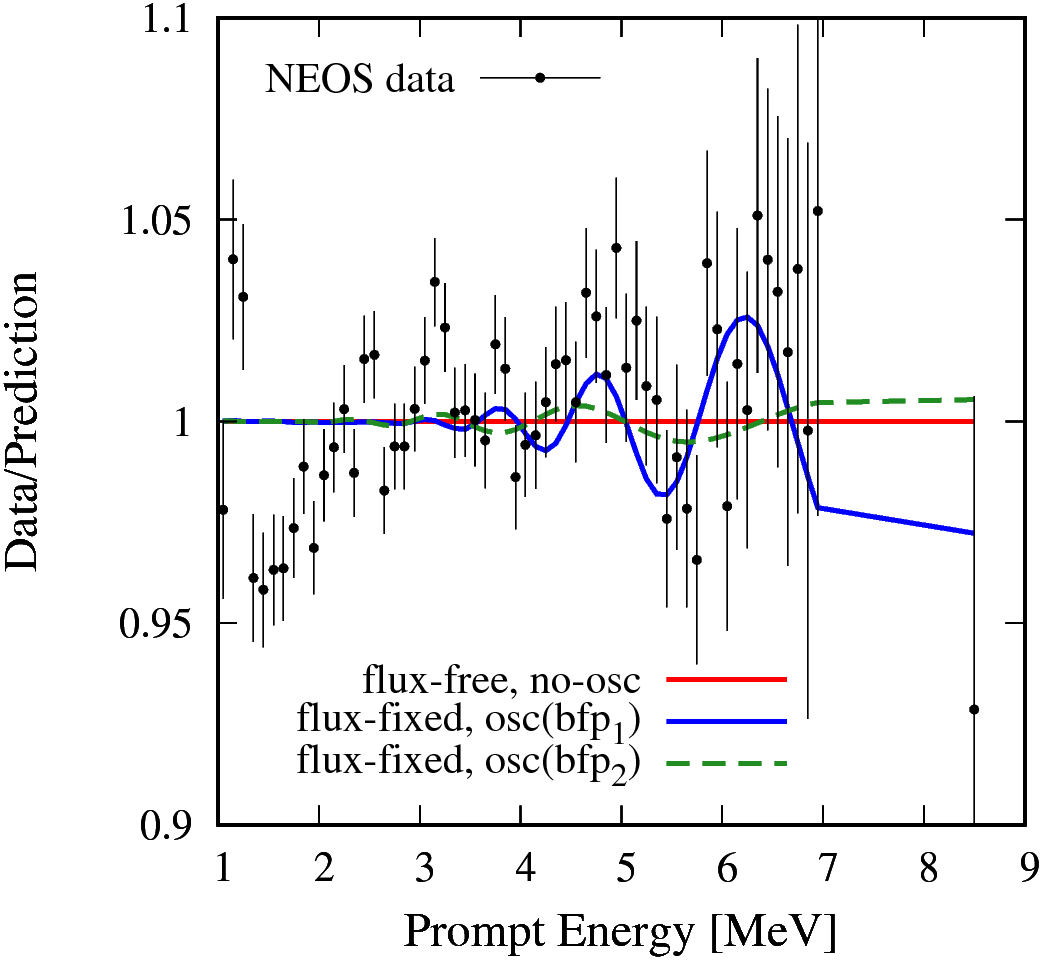}
  \hspace{1cm}
  \includegraphics[width=.45\textwidth]{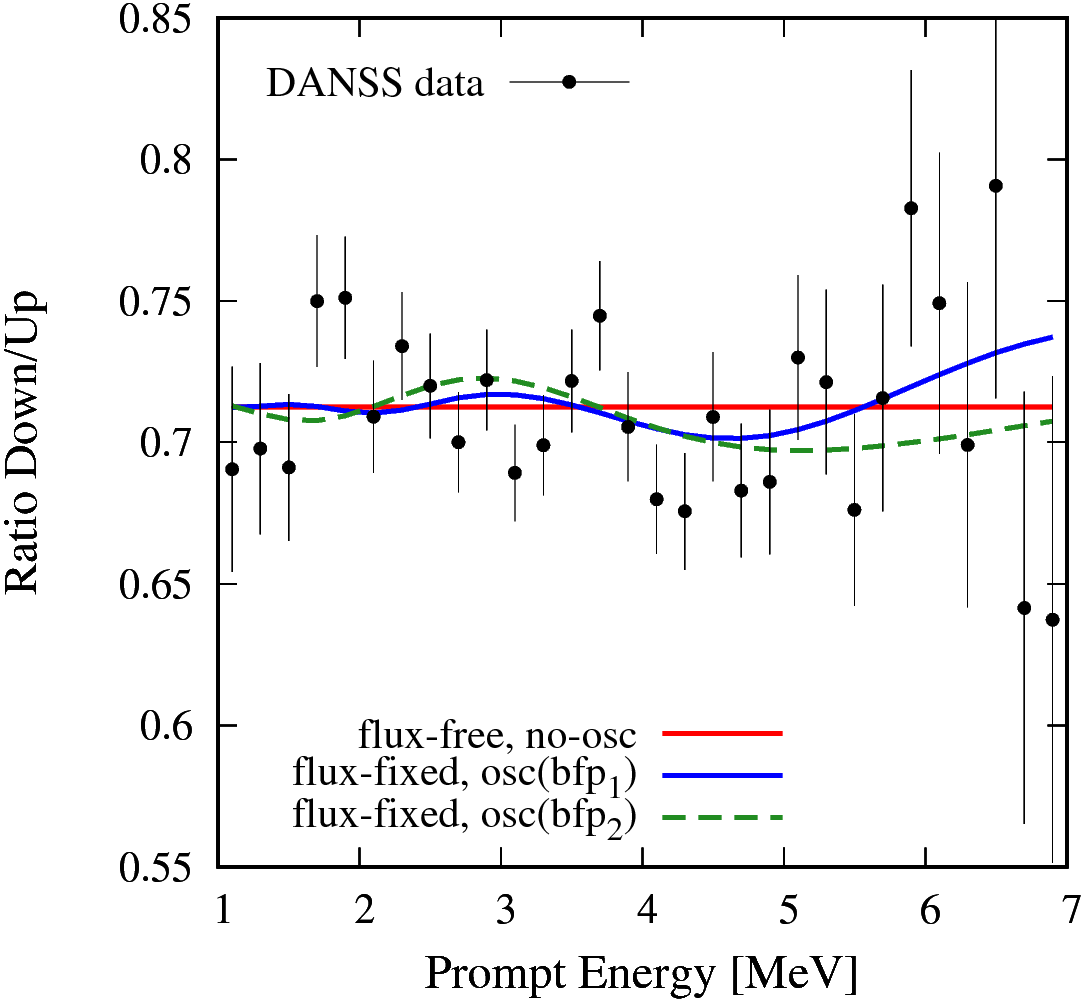}
  \mycaption{Spectral data for NEOS (left) and DANSS (right) compared
    to the flux-free no-oscillation prediction (red) and the
    predictions in case of flux-fixed + sterile neutrino oscillations,
    where bfp$_1$ (blue-solid curve) and bfp$_2$ (green-dashed curve)
    correspond to the best fit points from combined reactor data and
    global $\protect\parenbar{\nu}_e$ disappearance data,
    respectively, see \cref{tab:results}. Error bars correspond to
    statistical errors only. Details of the observables and
    predictions for the two experiments can be found in the
    appendix. The large distortions of the NEOS data below 2~MeV are
    within the systematic error band (not shown, see
    \cite{Ko:2016owz}).  }
  \label{fig:spectra}
\end{figure}

Both, NEOS and DANSS exclusion curves show strong wiggles in the RAA
region of 1~eV$^2 \lesssim \Delta m^2_{41} \lesssim 5$~eV$^2$. Those
features can be traced back to a slight oscillatory pattern of the
respective energy spectra, as shown in \cref{fig:spectra}, somewhat
more pronounced for NEOS (left panel) but also visible in DANSS (right
panel).  Indeed, the NEOS + Daya Bay analysis has a best fit point at
$\Delta m^2_{41} = 1.78$~eV$^2$ and $\sin^22\theta_{14} = 0.051$ with
\begin{equation}\label{eq:neos}
  \Delta\chi^2(\text{no osc.}) = 5.5  \qquad \text{(NEOS + Daya Bay spect)} \,.
\end{equation}
From the energy spectral data shown in \cref{fig:spectra} it is clear
that both, NEOS and DANSS data prefer oscillations for flux-fixed
compared to a constant re-scaling of fluxes. The effect is more
pronounced for the reactor-only best fit (blue-solid curve) but still
visible in the global disappearance analysis to be discussed in the
next section (green-dashed curves). Remarkably the wiggles in the
exclusion curves in \cref{fig:old-vs-new} from NEOS and DANSS partially match onto each other,
leaving large parts of the RAA region unconstrained. The red shaded
regions in \cref{fig:old-vs-new} include all data sets from
\cref{tab:reactor}. The fact that this region is pushed to larger
mixing angles compared to the ``old'' region is due to the Daya Bay
flux data, which prefer somewhat larger values of the mixing angle.

\begin{table}[t]
  \centering
  \begin{tabular}{llccrl}
    \hline\hline
    Data & Analysis & Best fit & $\chi^2_{\rm min}/\text{dof}$ & $\Delta\chi^2 (\text{no osc.})$ 
         & $p$-value$/\#\sigma$ \\
         &          & $(\sin^2 2\theta_{14}$, $\Delta m_{41}^2$) & & & (no osc.) \\
    \hline
    React-old & flux-fixed & (0.12, 1.72) & 52.1/68   &  9.4 \quad & \quad 0.0091$/$2.6$\sigma$\\
    React-old & flux-free  & (0.06, 0.46) & 51.6/66   &  2.8 \quad & \quad 0.25$/$1.2$\sigma$\\
    React-all & flux-fixed & (0.12, 2.99) & 196.0/236 & 11.3 \quad & \quad 0.0036$/$2.9$\sigma$ \\
    React-all & flux-free  & (0.04, 1.72) & 187.5/234 &  5.6 \quad & \quad 0.061$/$1.9$\sigma$\\
    Global    & flux-fixed & (0.06, 1.72) & 554.3/594 & 11.9 \quad & \quad 0.0026$/$3.0$\sigma$ \\
    Global    & flux-free  & (0.04, 1.72) & 545.2/592 &  7.0 \quad & \quad 0.031$/$2.2$\sigma$\\
    \hline\hline
  \end{tabular}
  \mycaption{Fit results for various data combinations (1st column)
    and assumptions about reactor fluxes (2nd column). Best fit points
    for $\Delta m_{41}^2$ are given in eV$^2$. The column
    ``$\Delta\chi^2 (\text{no osc.})$'' gives the difference in
    $\chi^2$ between $\theta_{14} = 0$ and the best fit point. The
    last column gives the $p$-value and the equivalent number of $\sigma$ obtained by evaluating the $\Delta \chi^2$
    for 2~dof.  The data samples are ``React-old'': reactor data sets
    predating the 2016 summer conferences, ``React-all'': combined
    analysis of all reactor neutrino experiments listed in
    \cref{tab:reactor}, ``Global'': combined
    $\protect\parenbar{\nu}_e$ disappearance data as discussed in
    \cref{sec:combined}.}
  \label{tab:results}
\end{table}

\begin{figure}[t]
  \centering \includegraphics[width=.67\textwidth]{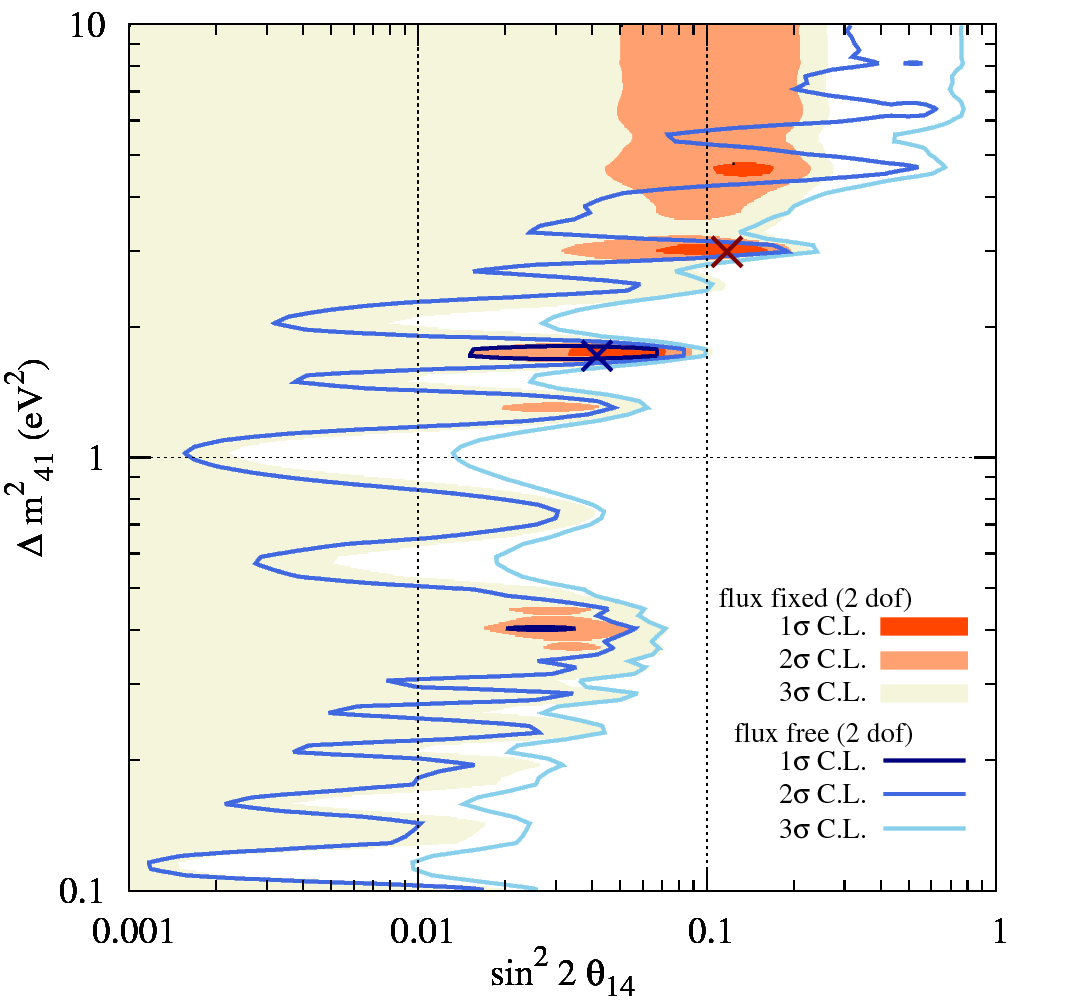}
  \mycaption{Allowed regions at $1,2,3\sigma$ (2~dof) confidence level
    for the combined analysis of all reactor neutrino experiments
    listed in \cref{tab:reactor}.  For the shaded regions we take the
    predicted fluxes and their quoted uncertainties according to Huber
    and Mueller~\cite{Mueller:2011nm, Huber:2011wv} at face value
    (``flux-fixed''), while for the unshaded contour lines, the fluxes
    from the four main fissible isotopes are allowed to vary freely
    (``flux-free''). The blue (red) cross indicates the best fit point
    for the flux-free (flux-fixed) analysis.  }
  \label{fig:react-all}
\end{figure}

\Cref{fig:react-all} shows the allowed regions at $1,2,3\sigma$
confidence level (2~dof) for the combined analysis of all reactor
neutrino experiments listed in \cref{tab:DBflux} and compares results
for the analyses with fixed and free fluxes.  Note that ``free
fluxes'' now includes also oscillations in addition to leaving fluxes
free in the fit, whereas before, it meant just rescaling of fluxes,
but $\theta_{14} = 0$.  In \cref{tab:results}, we summarize the best
fit points, the corresponding $\chi^2 / \text{dof}$ values, and the
$\Delta\chi^2$ between the best fit point and the no oscillation
hypothesis.  We observe that the significance of the RAA slightly
increases from a $p$-value for no-oscillations of 0.91\% ($2.6\sigma$)
for ``old'' data to 0.36\% ($2.9\sigma$) for combined reactor
data. Clearly for the flux-free analysis the significance for
oscillations decreases, but for the combined reactor data a hint for
oscillations remains even for flux-free ($p$-value of 6.1\%,
$1.9\sigma$), mostly driven by NEOS, cf.~\cref{eq:neos}.  Note that in
\cref{fig:react-all} the preferred regions from the flux-fixed
analysis are consistent with the flux-free exclusion limits.

\begin{table}[t]
  \centering
  \begin{tabular}{lccccc}
    \hline\hline
    React-all data & $(\sin^2 2\theta_{14}$, $\Delta m_{41}^2$) & $\xi_{235}$ & $\xi_{239}$ & $\xi_{241}$ & $\xi_{238}$ \\
    \hline
    flux-free, osc(bfp) & (0.04, 1.72) & 0.95 & 0.99 & 1.09 & 0.88\\
    flux-free, no-osc & -- & 0.93 & 0.96 & 1.09 & 0.87\\
    \hline\hline
  \end{tabular}
  \mycaption{Values of the fission fraction pull
      parameters for the flux free analysis at the oscillation best
      fit point and without oscillation.  These factors measure the
      relative rescaling of the IBD yields of each isotope with respect
      to the theoretical predictions, cf. \cref{eq:sigma-pred}.}
  \label{tab:pulls}
\end{table}

In \cref{tab:pulls} we provide the values of the pull
  parameters $\xi_i$, cf.~\cref{eq:sigma-pred}, obtained in the flux
  free analysis at the oscillation best fit point and for no
  oscillation.  In the latter case, the relative rescaling of the two
  main flux contributors, $\xi_{235}$ and $\xi_{239}$, qualitatively
  agree with the results in
  refs.~\cite{Giunti:2017yid,An:2017osx}\footnote{Note, however,
      that data sets differ.}, with $^{235}U$ being the
  main contributor to the flux deficit.  At the oscillation best fit point, the
  suppression for the $^{235}U$ and $^{239}P$ fluxes due to the pull parameters decrease because of
  the suppression from oscillations. But as in the no
  oscillation case, the larger suppression corresponds to $^{235}U$.
  Consistently, in the same way as the flux pulls decrease when
  including oscillations, the mixing angle at the best fit decrease
  when going from the flux fixed to the flux free analysis,
  cf.~\cref{tab:results}.

\begin{figure}[t]
  \centering
  \includegraphics[width=.48\textwidth]{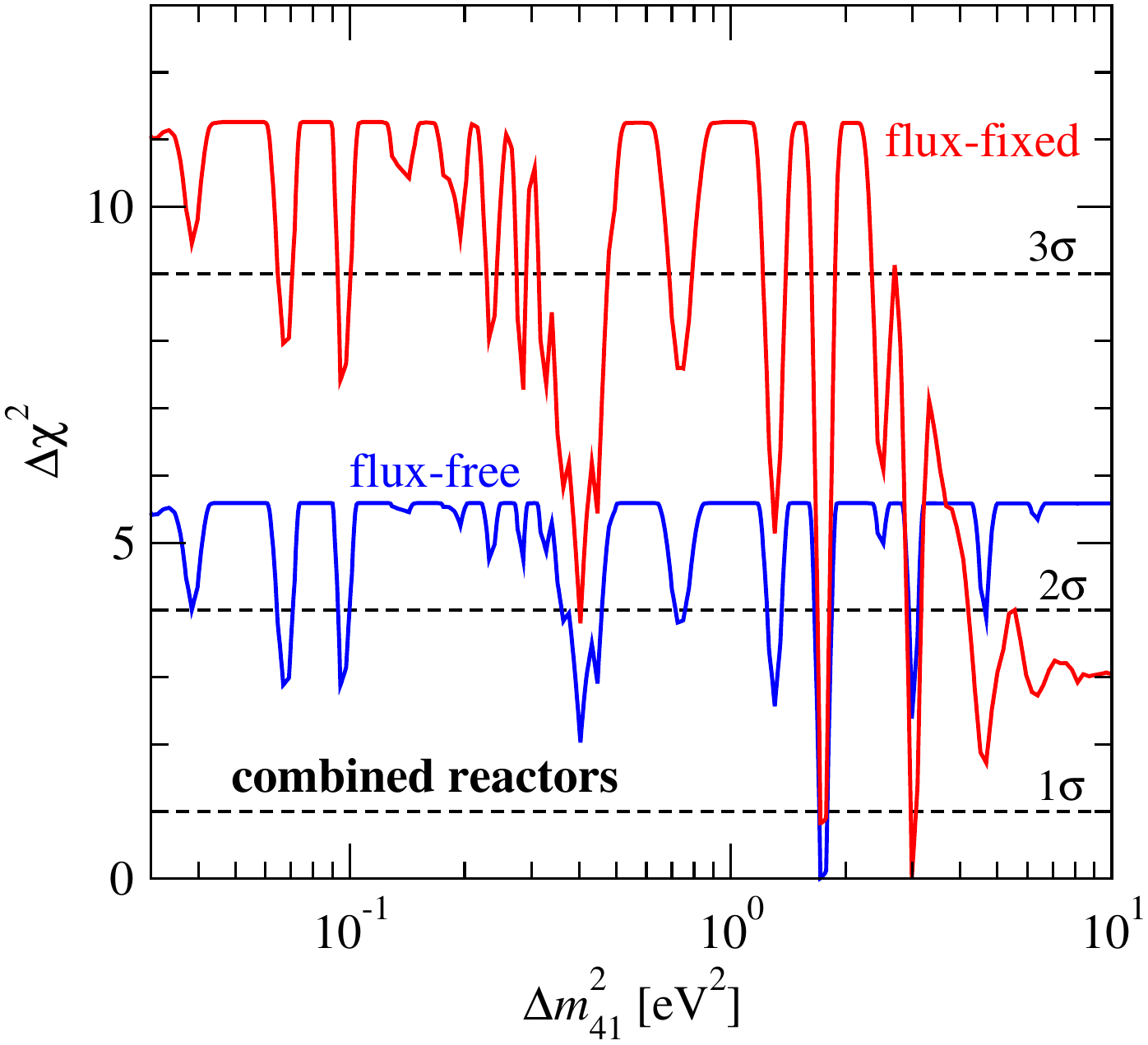}
  \includegraphics[width=.48\textwidth]{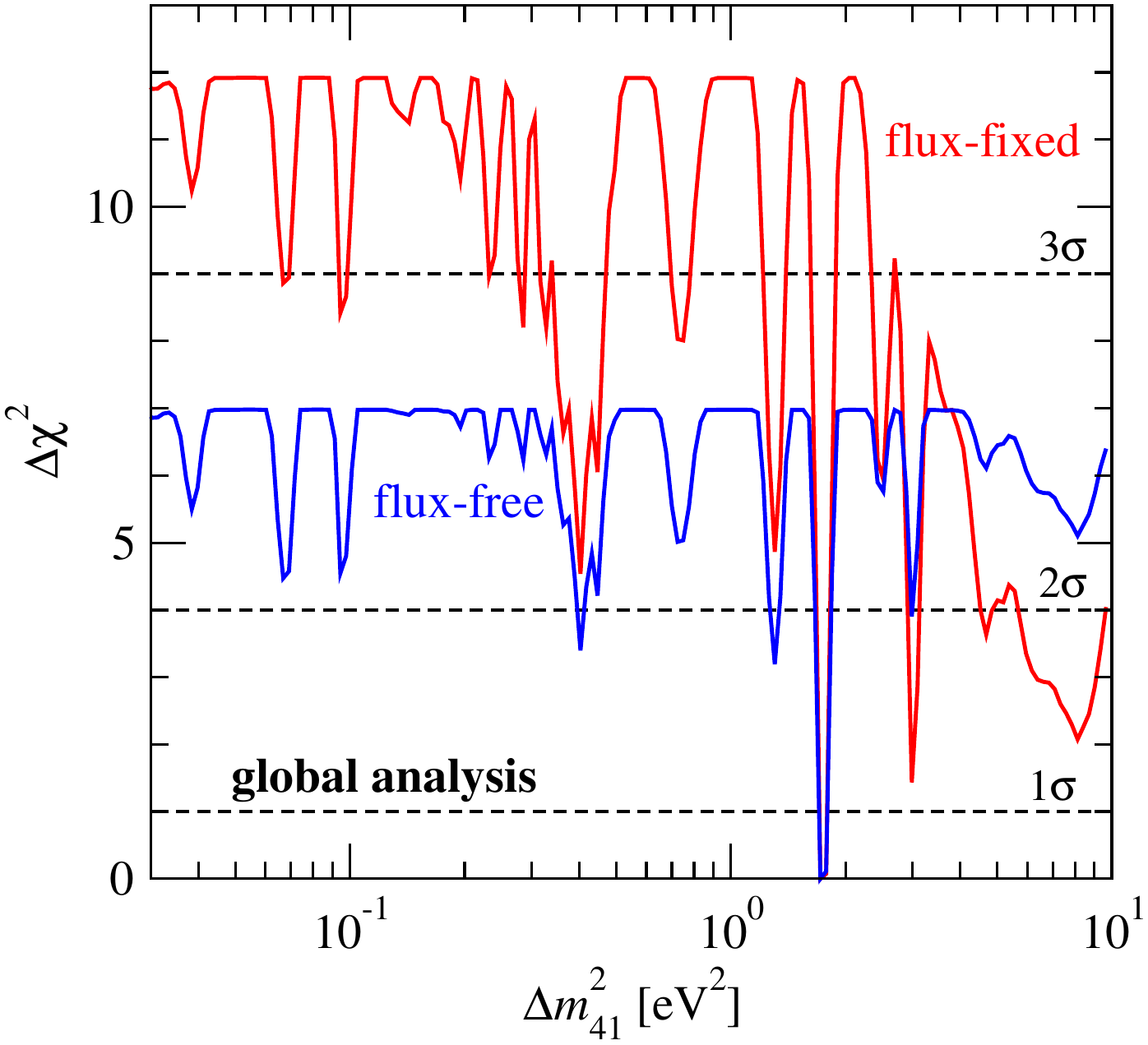}
  \mycaption{$\Delta\chi^2$ of the combined reactor (left) and global
    $\protect\parenbar{\nu}_e$ disappearance data (right) minimized
    with respect to all parameters except $\Delta m^2_{41}$, for
    free fluxes (blue curves) and fixed fluxes (red curves).}
  \label{fig:Dmq}
\end{figure}

In \cref{fig:Dmq} (left panel) we show the marginalized $\Delta\chi^2$
as a function of $\Delta m^2_{41}$ for both the fixed fluxes and free fluxes
analyses of the combined reactor data. We observe the two prominent
minima around 1.7 and 3~eV$^2$, both alowed at $1\sigma$ for
fixed fluxes. For free fluxes, we find the best fit
at 1.7~eV$^2$, with several other local minima below the $2\sigma$
threshold. Note that the maximal values of these curves correspond to
the $\Delta\chi^2$ for no oscillations as given in \cref{tab:results}.
The reason is that the no oscillation case can be obtained for
any $\Delta m^2_{41}$ when
minimizing $\chi^2$ with respect to the mixing angle.

We can also perform the same test as in \cref{sec:DBflux}, comparing
the two hypotheses flux-fixed + sterile neutrino versus flux-free
without sterile neutrino. From the numbers in \cref{tab:results} we
obtain for the test statistic $T$ defined in \cref{eq:T}:
\begin{align}\label{eq:Tobs-all}
  T_{\rm obs} = 2.9 & \qquad \text{(all reactors)} \,.
\end{align}
The spectral distortions observed in NEOS and DANSS prefer sterile
neutrino oscillations over flux rescaling and therefore the preference
for the flux-free fit obtained by Daya Bay flux data decreases. We
conclude that in light of the global data we cannot reject the sterile
neutrino hypothesis compared to the flux-free hypothesis. Let us
remark that due to spectral data, the sterile neutrino hypothesis is
now no longer a subset of the flux-free hypothesis (without
oscillations). Therefore the interpretation of above values for
$T_{\rm obs}$ in terms of $p$-values is not straight forward and we
limit our conclusion to qualitative statements relative to the result
obtained for Daya Bay flux data alone in \cref{eq:Tobs}.  Note that
when we use the same 10\% prior uncertainty on the subleading fluxes
\iso{U}{238} and \iso{Pu}{241} as in \cref{sec:DBflux} (instead of
20\% adopted in \cref{eq:Tobs-all}), the value for $T$ quoted in
\cref{eq:Tobs-all} decreases further to 2.1, to be compared to 6.3
obtained for Daya Bay flux alone, see \cref{eq:Tobs}.

Interestingly, for the old reactor data, oscillations are even
preferred over flux-free:
\begin{align}\label{eq:Tobs-old}
  T_{\rm obs} = -2.3 & \qquad \text{(old data)} \,.  
\end{align}
For this data set the best fit for oscillations is obtained at a
rather low value for the mass-squared difference around
0.4~eV$^2$. For this value, the observed rates at different baselines
can be fitted better than in the case of constant flux reduction,
which leads to a preferrence of oscillations. This result is in qualitative
agreement with ref.~\cite{Giunti:2017yid}, where further discussions
of this effect can be found. We note that in the global analysis of
all recent data, such low values of $\Delta m^2_{41}$ are disfavoured
at more than $3\sigma$ by spectral data, most importantly from Daya
Bay, cf.~\cref{fig:Dmq}, and therefore the flux-free hypothesis gives
still a slightly better fit than oscillations + the Huber \& Mueller
predictions.

Finally, let us note that DANSS result have not been published yet,
and our analysis is based on preliminary results presented in a
conference talk \cite{danss-moriond17}. We comment briefly on the
impact of those data on our result. Removing DANSS from the combined
analysis decreases the observed value for $T$ from 2.9 to 1.4, i.e.,
the sterile neutrino and the flux-free hypotheses become
statistically equivalent. Furthermore, the $\Delta\chi^2$ of
no-oscillations compared to the oscillation best fit point increases
by about 1 unit when removing DANSS, both for the flux-fixed and
flux-free analyses. While DANSS does show a weak preference for
oscillations, there is a slight tension between the best fit points
preferred with ($\Delta m^2_{41} \approx 3$~eV$^2$) and without
($\Delta m^2_{41} \approx 1.7$~eV$^2$) using DANSS,
c.f.\ \cref{fig:old-vs-new}. This leads to a slightly larger
preference in favour of oscillations when DANSS is left out. However,
quantitatively the impact is small and qualitatively the picture
remains robust, irrespective of using preliminary DANSS data or not.

\section{Global analysis of $\protect\parenbar{\nu}_e$ disappearance data}
\label{sec:combined}

\subsection{Non-reactor data}

\begin{table}
  \centering
  \catcode`!=\active\def!{\hphantom{0}}
  \begin{tabular}{lcclc} 
    \hline\hline 
    Experiment   & Ref. & \# Data & Comments & New? \\
    \hline
    \multicolumn{5}{l}{\bf Solar neutrino experiments} \\
    ~~Chlorine     & \cite{Cleveland:1998nv} & !!1 & rate & -- \\
    ~~GALLEX/GNO   & \cite{Kaether:2010ag} & !!2 & rates & -- \\
    ~~SAGE         & \cite{Abdurashitov:2009tn} & !!1 & rate & -- \\
    ~~Super-K phases 1--3 & \cite{Hosaka:2005um, Cravens:2008aa, Abe:2010hy}
                   & 119 & energy and zenith spectra & -- \\
    ~~Super-K phase 4 & \cite{sksol:nakano2016} & !46
                   & energy and day/night spectrum & \checkmark \\
    ~~SNO phases I--III & \cite{Aharmim:2007nv, Aharmim:2005gt, Aharmim:2008kc}
                   & !75 & energy and day/night spectra & -- \\
    ~~Borexino phase I  & \cite{Bellini:2011rx, Bellini:2008mr}
                   & !39 & low-energy and high-energy spectra & -- \\
    ~~Borexino phase II & \cite{Bellini:2014uqa}
                   & !42 & low-energy spectrum & \checkmark \\
    \hline
    \multicolumn{5}{l}{\bf Radioactive source experiments (gallium)} \\
    ~~GALLEX       & \cite{Hampel:1997fc, Kaether:2010ag} & !!2 && -- \\
    ~~SAGE         & \cite{Abdurashitov:1998ne, Abdurashitov:2005tb} & !!2 && -- \\
    \hline
    \multicolumn{5}{l}{\bf $\nu_e$ scattering on C-12
                     ($\nu_e + \iso{C}{12} \to \text{e}^- + \iso{N}{12}$)} \\
    ~~KARMEN       & \cite{Reichenbacher:2005nc, Armbruster:1998uk, Conrad:2011ce} & !26 && -- \\
    ~~LSND         & \cite{Auerbach:2001hz, Conrad:2011ce} & !!6 && -- \\
    \hline\hline
  \end{tabular}
  \mycaption{Experimental data which we combine with the reactor data
    from \cref{tab:reactor} in our global $\nu_e$/$\bar\nu_e$
    disappearance analysis. In the last column we indicate updates
    with respect to ref.~\cite{Kopp:2013vaa}. The total number of data
    points of non-reactor data is 361.}
  \label{tab:non-react}
\end{table}

In addition to the reactor neutrino data discussed before, there are other
experiments which are sensitive to $\protect\parenbar{\nu}_e$
disappearance and can therefore provide complementary information.
In this work, we consider in particular the data listed in
\cref{tab:non-react}:
\begin{itemize}
\item {\it Solar neutrino data.} We update our previous
  analysis~\cite{Kopp:2013vaa} by accounting for the 2055-day energy
  and day/night asymmetry spectrum from Super-Kamiokande
  phase~4~\cite{sksol:nakano2016}. We also include the new measurement of
  neutrinos from the proton-proton ($pp$) fusion chain in the Sun recently
  presented by Borexino~\cite{Bellini:2014uqa}. In addition, we make use of
  the total rates from the radiochemical experiments
  Chlorine~\cite{Cleveland:1998nv}, GALLEX/GNO~\cite{Kaether:2010ag}
  and SAGE~\cite{Abdurashitov:2009tn}, the electron scattering data binned
  in energy and zenith angle from all the previous Super-Kamiokande
  phases~\cite{Hosaka:2005um, Cravens:2008aa, Abe:2010hy}, the
  individual data sets from the three phases of
  SNO~\cite{Aharmim:2007nv, Aharmim:2005gt, Aharmim:2008kc}, and the
  Borexino phase-I data samples consisting of 740.7~days of low-energy
  data~\cite{Bellini:2011rx} and 246~live days of high-energy
  data~\cite{Bellini:2008mr}. Thus the solar neutrino data used in our
  analysis consists of 325 data points. Details of the implementation
  of the oscillation probabilities and relevant parameters can be
  found in Appendix~C of~\cite{Kopp:2013vaa}.

\item {\it Radioactive source experiments.} The calibration of Gallium solar
  neutrino experiments has been tested by deploying radioactive
  sources in the GALLEX~\cite{Hampel:1997fc, Kaether:2010ag} and
  SAGE~\cite{Abdurashitov:1998ne, Abdurashitov:2005tb} detectors.
  Both experiments have updergone two calibration campaigns: one with
  $^{37}\text{Ar}$ and one with $^{51}\text{Cr}$ in the case of SAGE,
  and both with $^{51}\text{Cr}$ in the case of GALLEX. All
  four measurements have reported an event rate about 10\%
  to 20\% lower than expected, a fact commonly known as the ``Gallium
  anomaly''. A re-evaluation~\cite{Frekers:2011zz} of the poorly-known
  contribution of $^{71}\text{Ge}$ excited states to the relevant
  $^{71}\text{Ga} \to {}^{71}\text{Ge}$ nuclear cross-section
  presented in~\cite{Bahcall:1997eg} has not settled the issue.  The
  deficit may be explained by the hypothesis of $\nu_e$ disappearance
  due to oscillations with a mass-squared difference at the eV$^2$
  scale, and is therefore a relevant ingredent of our study.  A
  detailed discussion of our implementation is provided in sec.~3.2
  of~\cite{Kopp:2013vaa}.

\item {\it $\nu_e$ scattering on $^{12}\text{C}$.} The
  LSND~\cite{Auerbach:2001hz} and KARMEN~\cite{Reichenbacher:2005nc,
    Armbruster:1998uk} experiments have measured the reaction $\nu_e +
  {}^{12}\text{C} \to \text{e}^- + {}^{12}\text{N}$, where the
  ${}^{12}$N decays back to ${}^{12}\text{C} + \text{e}^+ + \nu_e$
  with a lifetime of 15.9~ms. The experimental signature for this
  process, characterized by the observation of a prompt electron
  followed by a delayed positron, allows for precise identification
  of signal events and efficient rejection of backgrounds.  Both
  electron and positron energies are recorded, thus allowing to
  reconstruct the parent neutrino energy. No deviation from the
  no-oscillation hypothesis is observed in either detector, which
  results in a limit on the sterile neutrino admixture to
  $\nu_e$~\cite{Conrad:2011ce}. Details of our implementation of LSND
  and KARMEN results on ${}^{12}\text{C}$ scattering are provided in
  Appendix~E.1 of~\cite{Kopp:2013vaa}.
\end{itemize}

In our analysis, correlations among the various experimental
results within each of the three classes of data listed above (solar,
radioactive source, scattering on $^{12}\text{C}$) are properly taken
into account, whereas correlations between different classes are
neglected. In principle, the GALLEX and SAGE experiments contribute
both to the solar neutrino analysis and to the Gallium anomaly, thus
introducing a correlation among these two sets. However, we
have verified that the solar neutrino rate in GALLEX and SAGE is
completely dominated by the ground-state cross-section, which has a
small error. Conversely, the main source of
uncertainty affecting the Gallium anomaly comes from the two excited
levels GT175 and GT500 (see~\cite{Kopp:2013vaa} for details), whose
contribution to the solar neutrino interaction rate is only at the
percent level. Therefore, a proper treatment of the correlations
between the Gallium anomaly and solar neutrino data, despite
introducing a non-trival complication, would add very little to the
results of our study.

\subsection{Results}

\begin{figure}[t]
  \centering
  \includegraphics[width=.49\textwidth]{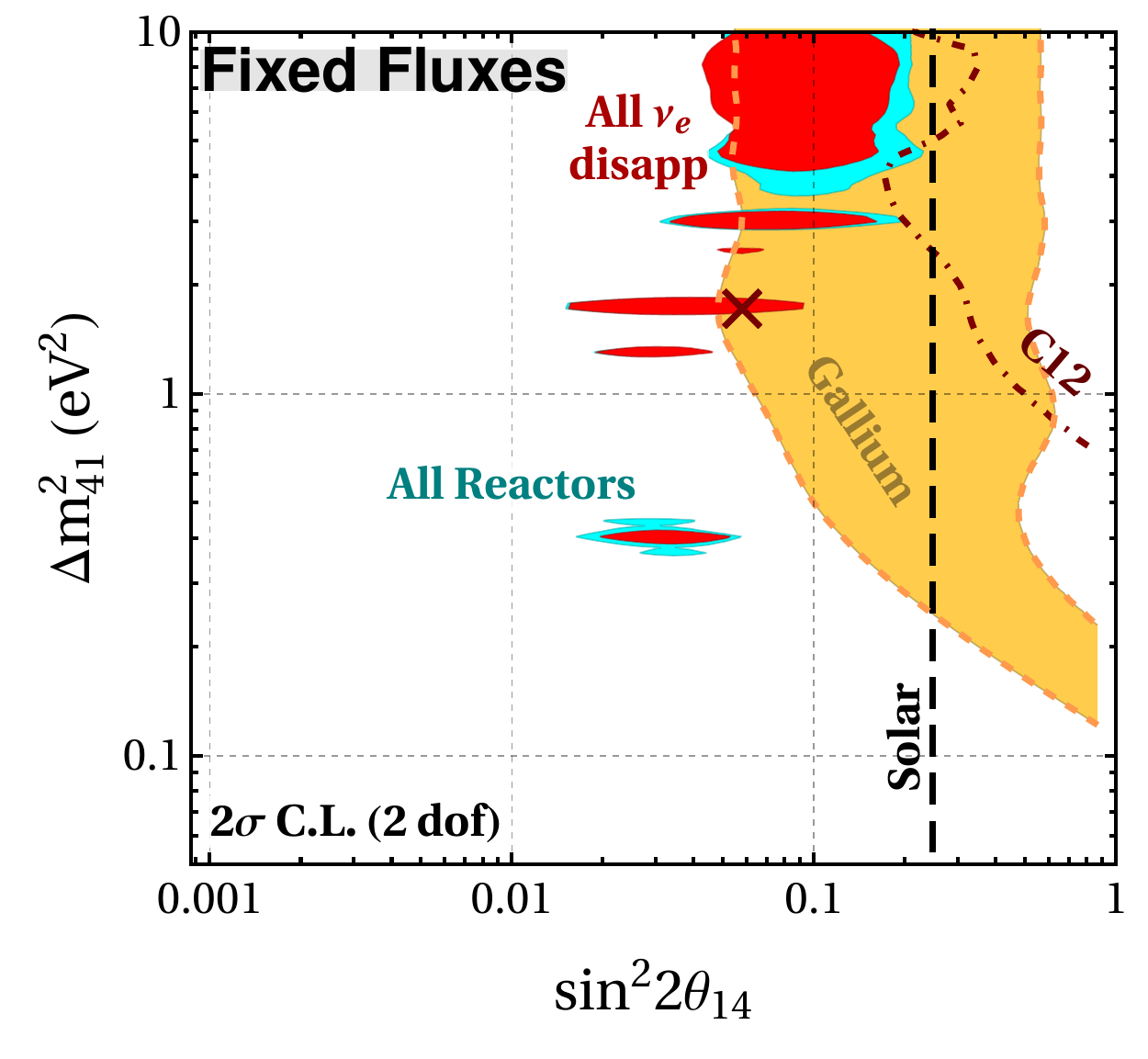}
  \includegraphics[width=.49\textwidth]{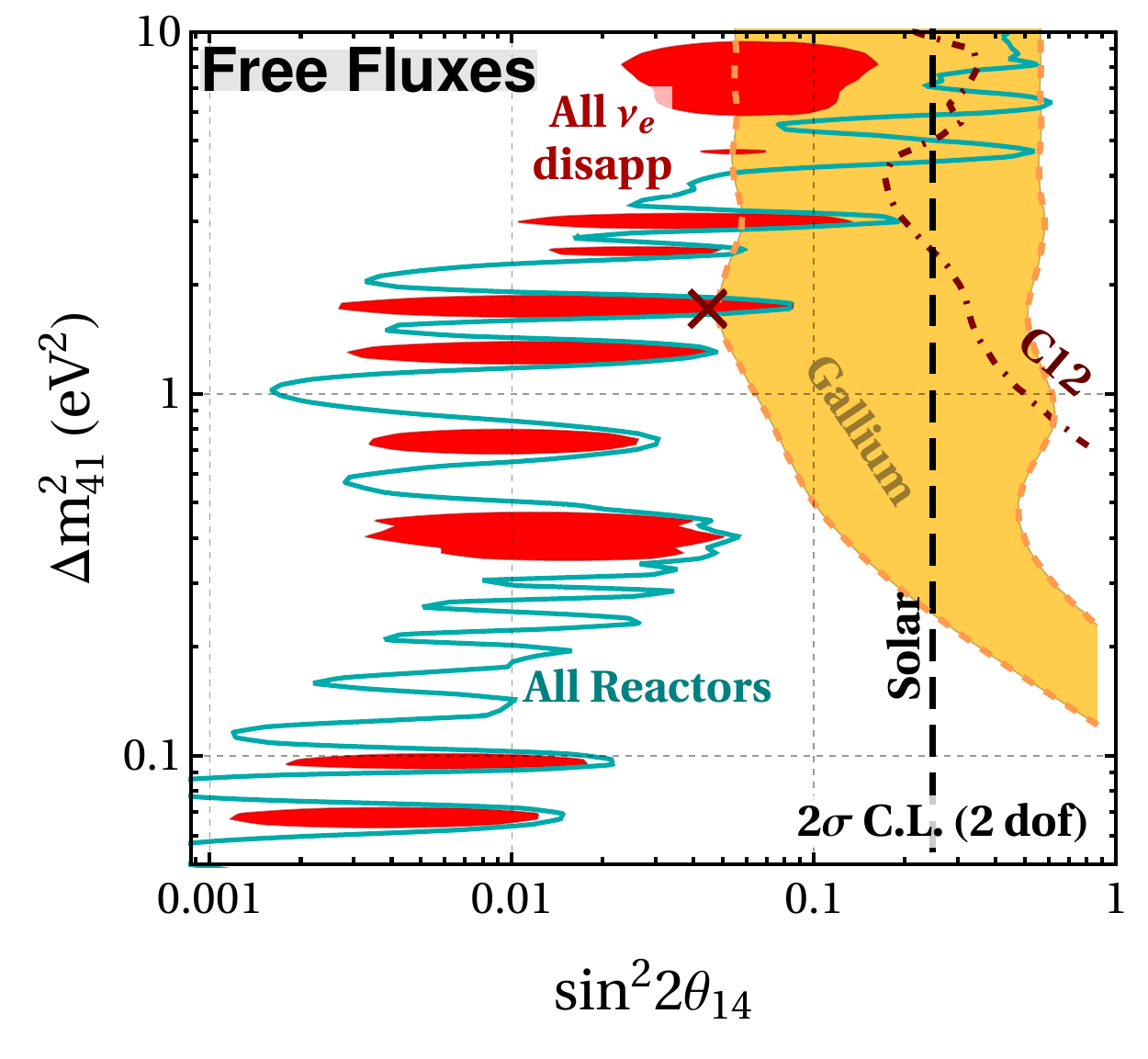}
  \mycaption{Allowed parameter regions at $2\sigma$ (2~dof) for the
    ``fixed fluxes'' (left) and ``free fluxes'' (right) analyses, for
    all $\protect\parenbar{\nu}_e$ disappearance experiments (red
    shaded regions). The global best fit point is marked with a cross.
    In addition we show the regions or bounds
    obtained from combined reactor data, the radioactive source
    experiments, $\nu_e$ scattering on \iso{C}{12}, and solar
    data. }
  \label{fig:all_nu_e_disapp}
\end{figure}

The results of our global analysis of all $\protect\parenbar{\nu}_e$
disappearance experiments are shown in
\cref{fig:all_nu_e_disapp}
for the ``fixed fluxes'' and
``free fluxes'' analyses.  $\Delta\chi^2$ profiles as a
function of $\Delta m^2_{41}$ are shown in the right panel of
\cref{fig:Dmq}.
Best-fit points and $\chi^2$ values are reported in the last two rows of
\cref{tab:results}. 
We observe that, both for free fluxes and for fixed fluxes,
the combined fit is largely dominated by reactor neutrino data.
The total number of data points in this analysis
is 600, and the oscillation fit includes the six parameters $\Delta
m^2_{41}$ and the mixing angles
$\theta_{12},\theta_{13},\theta_{14},\theta_{24},\theta_{34}$; the
other mass-squared differences and $\theta_{23}$ are fixed to their
3-flavour best fit points. Although we do take into account the two
complex phases on which solar oscillation probabilities formally
depend, their impact on the $\chi^2$ is negligible and we do not count
them as degree of freedom in the fit, see appendices of
ref.~\cite{Kopp:2013vaa} for a discussion of complex phases.

For what concerns solar neutrino data, the mass-squared difference
$\Delta m^2_{41}$ implied by the reactor anomaly is virtually infinite
in the calculation of the $P_{ee}$ survival probability, hence its
specific value is not constrained by solar experiments. The bound on
$\theta_{14}$ is mainly driven by the good agreement between the
theoretical expectation of the $^{8}\text{B}$ neutrino flux, which is
predicted by the Standard Solar Model, and its precise determination
in high-energy solar experiments. This includes direct measurements
(through neutral current interactions in SNO) and indirect measurements (through
the combination of charged current and elastic scattering data in SNO
and SK). The inclusion of a sterile neutrino admixture with $\nu_e$
implies an overall reduction of the flux of active neutrinos at the
detector, thus spoiling such agreement. This results in an upper bound
on $\theta_{14}$, which in the case of the ``fixed flux'' analysis is
fully compatible with the entire region allowed by reactor data, thus
adding little to the global analysis. In the ``free fluxes'' case, solar
data help restricting $\theta_{14}$ at $\Delta
m^2_{41} \gtrsim 4$~eV$^2$, where reactor experiments lose
sensitivity because the oscillation length becomes very short,
implying a uniform suppression of the reactor neutrino flux
in all reactor experiments, but no spectral features. Such a
uniform suppression cannot be disentangled from a
rescaling of the flux normalization. Similar arguments also apply to
the LSND \& KARMEN data on $^{12}\text{C}$, which show no deviation
from the standard oscillation scenario and therefore impose an upper
bound on $\theta_{14}$ in the large $\Delta m^2_{41}$ region.

The situation regarding the Gallium anomaly is somewhat
different. As already explained, the GALLEX and SAGE experiments
observe a deficit which can be interpreted in terms of sterile
neutrino oscillations. However, its $2\sigma$ allowed region shows
little overlap with the reactor region, except for a small area at
large $\Delta m^2_{41}$. In general, Gallium data favor a larger value
of $\theta_{14}$ than reactor data. It should be noted, however, that
while the \emph{lower} bound on $\theta_{14}$ from GALLEX and SAGE is
rather weak, with the no-oscillation value $\theta_{14} = 0$
disfavoured only by $\Delta\chi^2 = 8.72$ with respect to the best-fit
point (see sec.~3.2 of Ref.~\cite{Kopp:2013vaa}), the \emph{upper} bound on
$\theta_{14}$ from reactor data is pretty strong for $\Delta m^2_{41}
\lesssim 5$~eV$^2$. Therefore, a combination of reactor and gallium
data naturally favors the reactor region, rather than the GALLEX and
SAGE one, so that the net contribution of gallium data is vastly
reduced.

Indeed, as can be seen also in \cref{tab:results} and \cref{fig:Dmq},
the results of the global analysis differ little from those of the
reactor-only one, with a very similar $\Delta\chi^2$ for no
oscillations in the ``fixed fluxes'' analysis of 11.3 versus 11.9
($p$-values of 0.36\% versus 0.26\%). For free fluxes, the impact of
Gallium data is somewhat larger, increasing the $\Delta\chi^2$ of no
oscillations from 5.6 for reactors to 7.0 for the global data
($p$-values of 6.1\% versus 3.1\%). This corresponds to a hint in
favour of oscillations at $2.2\sigma$, resulting in closed regions at
the $2\sigma$ level for the flux free analysis, visible in
\cref{fig:all_nu_e_disapp} (right panel).

The test statistic $T$ defined in \cref{eq:T} for
discriminating between the flux-fixed + oscillations versus flux-free
+ no oscillations decreases from 2.9 (reactor-only data) to 2.1 for
the global disappearance data.

\section{Discussion and Conclusions}
\label{sec:discussion}

In this paper we have investigated the status of the sterile neutrino
hypothesis in the context of the global data on
$\protect\parenbar{\nu}_e$ disappearance, in the light of new results
from reactor neutrino experiments. In particular, we have considered
the impact of first results from the NEOS and DANSS short-baseline
reactor experiments, as well as the recent determination of the
inverse-beta decay rate induced by neutrinos from different fission
isotopes by Daya Bay. In our reactor data analysis we have taken into
account the disagreement of data and predictions in the spectral shape
(``5~MeV bump'') by using only relative spectra at different
baselines.

We confirm that Daya Bay flux measurements alone favour the hypothesis
that the source of the reactor anomaly is the flux of
\iso{U}{235}\ over the hypothesis of sterile neutrino
oscillations. However, the sterile neutrino hypothesis also provides a
good description of the data ($p$-value of 18\%) and hence cannot be
excluded. Therefore we combine this Daya Bay result with the remaining
data from reactor experiments assuming the presence of a sterile
neutrino. For the global reactor data, actually, the preference for
re-scaling the \iso{U}{235}\ flux over oscillations is reduced
compared to the Daya Bay flux data alone, see
\cref{eq:Tobs,eq:Tobs-all}, and the sterile neutrino hypothesis +
Huber \& Mueller flux predictions cannot be rejected. The main reason
for this are features in the energy spectra reported by the DANSS and
NEOS experiments, which prefer oscillations over a rescaling of fluxes.

Since the sterile neutrino hypothesis cannot be excluded, we present
updated determinations of oscillation parameters. The shape of the
exclusion curves from DANSS and NEOS spectral data leaves allowed
parameter space consistent with each other and with the remaining
reactor neutrino data. The combined analysis leads to islands in the
allowed parameter space around $(\Delta m^2_{41},\sin^22\theta_{14}) \sim
(1.7\,{\rm eV}^2,0.04)$ and $(3\,{\rm eV}^2,0.1)$,
and the significance of the sterile neutrino explanation of the
reactor anomaly remains slightly below $3\sigma$ (the no-oscillation
hypothesis has a $p$-value relative to the best fit point of
0.36\%). We have also preformed an oscillation fit leaving the neutrino
fluxes from the four main fission isotopes completely free. Although
the significance of the anomaly decreases, a hint for
oscillations remains at the $1.9\sigma$ level and the exclusion curves on the oscillation
parameters are consistent with the best fit regions obtained in the
analysis with fixed fluxes.

Finally, we have provided updated results from a global fit to
$\protect\parenbar{\nu}_e$ disappearance data, including the Gallium
anomaly and constraints from solar neutrinos and from
$\nu_e$--$\iso{C}{12}$ scattering in LSND and KARMEN.  The results of
the global analysis are largely consistent with the reactor-only fit,
and the indications for sterile neutrino oscillations remain at a
significance close to $3\sigma$ ($2\sigma$) with respect to no
oscillations in the case of flux-fixed (flux-free) analyses.

In conclusion, present data on $\protect\parenbar{\nu}_e$
disappearance is still consistent with sterile neutrino oscillations at the eV
scale with modest significance. To definitely clarify the question raised in the title, more
data is needed, which can be expected in the near future from new
short-baseline reactor experiments as well as radioactive source
experiments, see e.g., ref.~\cite{Gariazzo:2017fdh} for references and
a review of sensitivities of upcoming experiments.

\section*{Acknowledgments}

We thank Bryce Littlejohn for useful comments on the manuscript.
The work of JK and MD has been supported by the German
Research Foundation (DFG) under Grant Nos.\ \mbox{KO~4820/1--1}, FOR~2239,
EXC-1098 (PRISMA) and by the European Research Council (ERC) under the European
Union's Horizon 2020 research and innovation programme (grant agreement No.\
637506, ``$\nu$Directions'').
AHC, MM and TS are supported by the European Union’s Horizon
2020 research and innovation programme under the Marie
Sklodowska-Curie grant agreement No 674896 (Elusives).
The work of MM is supported also by 
the EU Network INVISIBLES-PLUS (H2020-MSCA-RISE-2015-690575),
MINECO/FEDER-UE grants FPA2015-65929-P and FPA2016-78645-P, and 
the ``Severo Ochoa'' program grant SEV-2016-0597 of IFT.

\appendix

\section{Technical details on our analyses}

\subsection{Daya Bay sterile neutrino fit}
\label{app:DBsterile}

Using the data from the different Daya Bay experimental halls and
in the different energy bins, a $\chi^2$ depending on $\theta_{13}$,
$\theta_{14}$, $\Delta m^2_{41}$ and pull parameters is computed as
follows ($\Delta m^2_{31}$, $\Delta m^2_{21}$, and $\theta_{12}$ are
included but kept fix at their 3-flavour best fit values):
\begin{align}
  \chi^2(\theta_{13},\theta_{14},\Delta m^2_{41}, \textbf{p})
  &= \sum_{i=1}^{35} \frac{1}{\sigma^\text{stat,31}_i}
     \bigg[ \frac{O^3_i-B^3_i}{O^1_i-B^1_i}
          - \frac{N^3_i}{N^1_i}(\theta_{13},\theta_{14},\Delta m^2_{41}, \textbf{p}) \bigg]^2\nonumber
  \\
  &+ \sum_{i=1}^{35} \frac{1}{\sigma^\text{stat,21}_i}
     \bigg[ \frac{O^2_i-B^2_i}{O^1_i-B^1_i}
          - \frac{N^2_i}{N^1_i}(\theta_{13},\theta_{14},\Delta m^2_{41}, \textbf{p}) \bigg]^2
   + \textbf{p}^T V_{\textbf{p}}^{-1} \textbf{p} \,.
\end{align}
Here, $O^{H}_i$ are the observed number of events in experimental hall $H$ and
energy bin $i$, and $B^{H}_i$ are the corresponding predicted background. The
measured event rates and background predictions are taken from the
supplementary material of ref.~\cite{An:2016ses}. $N^{H}_i$ are the predicted
event numbers in experimental hall $H$ and bin $i$, see below.
$\sigma^{\text{stat},H H'}_i$ are the statistical errors of the ratios
$(O^{H}_i - B^{H}_i)/(O^{H'}_i - B^{H'}_i)$. Finally, $\textbf{p}$ is the
vector of nuisance parameters accounting for the systematic uncertainties, and
$V_\textbf{p}$ is the corresponding covariance matrix. It includes the
uncertainties in the relative energy scale and the detection efficiency as well
as their correlation, which can be found in table~VIII of
ref.~\cite{An:2016ses}. Since we are using bin-by-bin ratios between detectors
at different baselines, errors in the flux predictions and in the inverse beta
decay cross sections will mostly cancel.  The minimization over the pull
parameters is done by linearizing the dependence of $N^H_i$ on $\textbf{p}$ and then
solving a linear system of equations.

Since each experimental hall in Daya Bay houses several detectors, $N^H_i$
is obtained by summing the contributions from all detectors in hall $H$.
The predicted number of events in an individual detector $d$ and energy bin $i$ is
\begin{align}
  N_i^d &= A^H \sum_r \sum_\text{iso} \frac{\epsilon^d}{L^2_{rd}}
           \int_{E^\text{rec}_i}^{E^\text{rec}_{i+1}} \! dE^\text{rec} \int_0^\infty \! dE_{\nu} \,
           \sigma(E_{\nu}) \, f^\text{iso} \phi^\text{iso}(E_{\nu})
           P^{rd}_{\bar\nu_e \to \bar\nu_e}(E_{\nu}) R(E^\text{rec},E_{\nu}) \,,
  \label{eq:Nid}
\end{align}
where
\begin{itemize}
  \item the indices $i$, $r$, $d$ and $iso$ refer to energy bins, reactors,
    detectors, and fissible isotopes, respectively.
  \item $\epsilon^d$ is the detector efficiency, taken from table~VI in
    ref.~\cite{An:2016ses}. We consider the efficiencies $\epsilon_\mu$
    and $\epsilon_m$ (corresponding to loss of events from the muon veto and
    multiplicity veto, respectively) as well as the variation in the number of
    target protons in each detector, $\Delta N_p$.
  \item $L_{rd}$ is the baseline between reactor $r$ and detector $d$.
  \item $E_\nu$ and $E^\text{rec}$ are the true neutrino energy and the
    energy reconstructed by the detector, respectively. The detector response
    function $R(E^\text{rec},E_{\nu})$, taken from the supplementary material of
    ref.~\cite{An:2016ses}, describes the mapping between $E_\nu$ and $E^\text{rec}$.
  \item $\sigma(E_{\nu})$ is the inverse beta decay cross section~\cite{Vogel:1999zy}.
  \item $\phi^\text{iso}(E_{\nu})$ are the flux predictions from refs.~\cite{Mueller:2011nm,
    Huber:2011wv}, and $f^\text{iso}$ are the fission fractions. For each isotope,
    $f^\text{iso}$ is computed as the average of the fission fractions in table~9
    of ref.~\cite{An:2016srz}.
  \item $P^{rd}_{\bar\nu_e \to \bar\nu_e}(E_{\nu})$ is the oscillation probability.
  \item $A^H$ is a normalization factor, which is fixed by requiring the total
    predicted number of events in hall $H$ without oscillations to match the
    corresponding number given in the supplementary material of ref.~\cite{An:2016ses}.
\end{itemize}

\subsection{NEOS}
\label{app:neos}

Our fit to NEOS data is based on fig. 3(c) of ref.~\cite{Ko:2016owz}, where the
data are presented as ratios between observed event rates in NEOS and a
prediction based on the unfolded Daya Bay anti-neutrino spectrum from
ref.~\cite{An:2016srz}.  We adopt the following $\chi^2$ function:
\begin{align}
  \chi^2(\theta_{14}, \Delta m^2_{41}) &=
    \sum_{i,j=1}^{60} \Big[ O^N_i - P^N_i(\theta_{14},\Delta m^2_{41}) \Big]
          V^{-1}_{ij} \Big[ O^N_j - P^N_j(\theta_{14},\Delta m^2_{41}) \Big] \,.
  \label{eq:neos-chisq}
\end{align}
Here, $O^N_i$ is the NEOS data point in energy bin $i$, and $P^N_i$ is
the theoretical prediction.  To obtain the latter, we have to account for
the fact that the unfolded Daya Bay spectrum is based on the assumption of
three-flavour oscillations.  We therefore have to unfold three-flavour oscillations
(which are, however, small in Daya Bay and negligible in NEOS)
and fold in four-flavour oscillations:
\begin{align}
  P^N_i = \frac{P^\text{NEOS}_{4 \nu,i}}{P^\text{NEOS}_{3 \nu,i}}
          \frac{P^\text{DB}_{3 \nu,i}} {P^\text{DB}_{4 \nu,i} } \,,
\end{align}
where $P^\text{Exp}_{n\nu}$ are the predicted event numbers in bin $i$ for
experiment $\text{Exp}=\text{NEOS},\ \text{DB}$ in the $n\nu$ neutrino
framework. The latter are obtained based on the Huber--Mueller fluxes
\cite{Mueller:2011nm, Huber:2011wv}. The
covariance matrix $V_{ij}$ in \cref{eq:neos-chisq} includes statistical errors
extracted from fig.~3(c) in ref.~\cite{Ko:2016owz} as well as the covariance
matrix for the Daya Bay flux determination provided in ref.~\cite{An:2016srz}.

For the Daya Bay predictions we take into account an average of the near
detectors (EH1 and EH2) as used for the Daya Bay unfolded spectrum in
\cite{An:2016srz} and they are calculated as in \cref{app:DBsterile}.  The
number of events per bin in NEOS is computed in analogy to \cref{eq:Nid}. Since
the NEOS detector is very close to the source, we also take into account the
finite sizes of the reactor core and of the detector by averaging the
oscillation probability weighted by $1/L^2$ over $L=(24 \pm 1.5)$~m.  Since no
response function $R(E^\text{rec},E_{\nu})$ is provided by the NEOS
collaboration we adopt the model proposed in ref.~\cite{Huber:2016xis}
consisting of a a Gaussian for $E^\text{rec} > E_p$ ($E_p = E_{\nu} - 0.8$~MeV)
and a rescaled Gaussian plus a constant value for $E^\text{rec} < E_p$ to account
for photons or positrons escaping the detector. In order to reproduce the NEOS
spectrum from figure~3(b) in \cite{Ko:2016owz}, we assume energy scale
non-linearity effects based on the information on non-linearity provided by
Daya Bay in the supplementary material of ref.~\cite{An:2016ses}.

\subsection{DANSS}
\label{app:danss}

For the DANSS experiment, we use the preliminary data shown on slide~10
of ref.~\cite{danss-moriond17}.
The data are given as ratios of observed event numbers between the two
detector positions at $L=12.7$~m (down) and $L=10.7$~m (up) from the center
of the reactor core. The data are divided into 30~energy bins
of equal width, ranging from $E_p = 1.0$~MeV to $E_p = 7.0$~MeV. Here,
$E_p$ is the kinetic energy of the outgoing positron in inverse beta decay
$\bar\nu_e + p \to n + e^+$.  The $\chi^2$ for DANSS is
\begin{align}
  \chi^2(\theta_{14}, \Delta m^2_{41}) &=
    \sum_{i,j=1}^{30} \Big[ O_i - P_i(\theta_{14},\Delta m^2_{41}) \Big]
          V^{-1}_{ij} \Big[ O_j - P_j(\theta_{14},\Delta m^2_{41}) \Big] \,,
  \label{eq:danss-chisq}
\end{align}
where the predicted down/up ratios $P_i$ are computed as ratios of
oscillation probabilities, weighted by the geometric factor $1/L^2$.
To account for the size and geometry of the detector and the reactor,
we average the oscillation probabilities (divided by $L^2$) over
$L = L_0 \pm 4.0\,\text{m}$. Here, $L_0$ is taken to be $12.85$\,m
for the lower detector position and $10.9$\,m for the upper one.
These numbers are slightly larger than the distances between the
center of the reactor core and the center of the detector to account
for the on average non-zero horizontal distance between the production
and detection vertices.  The energy resoluton of DANSS is modeled
as a Gaussian with a width given by fig.~5 of ref.~\cite{Danilov:2014vra}.
The covariance matrix $V_{ij}$ for DANSS includes only statistical
uncertainties and a 2\% systematic uncertainty on the down/up ratios.

\bibliographystyle{JHEP_improved}
\bibliography{./refs}

\providecommand{\href}[2]{#2}\begingroup\raggedright\begin{thebibliography}{10}

\bibitem{Mueller:2011nm}
T.~A. Mueller et~al., \href{http://dx.doi.org/10.1103/PhysRevC.83.054615}{{\it
  {Improved Predictions of Reactor Antineutrino Spectra}}, } {\em Phys. Rev.}
  {\bf C83} (2011) 054615, [\href{http://arxiv.org/abs/1101.2663}{{\tt
  1101.2663}}].

\bibitem{Huber:2011wv}
P.~Huber, \href{http://dx.doi.org/10.1103/PhysRevC.84.024617}{{\it {On the
  determination of anti-neutrino spectra from nuclear reactors}}, } {\em Phys.
  Rev.} {\bf C84} (2011) 024617, [\href{http://arxiv.org/abs/1106.0687}{{\tt
  1106.0687}}].

\bibitem{Schreckenbach:1985ep}
K.~Schreckenbach, G.~Colvin, W.~Gelletly, and F.~{Von Feilitzsch},
  \href{http://dx.doi.org/10.1016/0370-2693(85)91337-1}{{\it {Determination of
  the anti-neutrino spectrum from U235 thermal neutron fission products up to
  9.5 MeV}}, } {\em Phys.Lett.} {\bf B160} (1985) 325--330.

\bibitem{Hahn:1989zr}
A.~Hahn, K.~Schreckenbach, G.~Colvin, B.~Krusche, W.~Gelletly, et~al.,
  \href{http://dx.doi.org/10.1016/0370-2693(89)91598-0}{{\it {Anti-neutrino
  spectra from Pu241 and Pu239 thermal neutron fission products}}, } {\em
  Phys.Lett.} {\bf B218} (1989) 365--368.

\bibitem{VonFeilitzsch:1982jw}
F.~{Von Feilitzsch}, A.~Hahn, and K.~Schreckenbach,
  \href{http://dx.doi.org/10.1016/0370-2693(82)90622-0}{{\it {Experimental beta
  spectra from Pu239 and U235 thermal neutron fission products and their
  correlated anti-neutrino spectra}}, } {\em Phys.Lett.} {\bf B118} (1982)
  162--166.

\bibitem{Vogel:1980bk}
P.~Vogel, G.~Schenter, F.~Mann, and R.~Schenter,
  \href{http://dx.doi.org/10.1103/PhysRevC.24.1543}{{\it {Reactor anti-neutrino
  spectra and their application to anti-neutrino induced reactions. 2.}}, }
  {\em Phys.Rev.} {\bf C24} (1981) 1543--1553.

\bibitem{Mention:2011rk}
G.~Mention et~al., \href{http://dx.doi.org/10.1103/PhysRevD.83.073006}{{\it
  {The Reactor Antineutrino Anomaly}}, } {\em Phys. Rev.} {\bf D83} (2011)
  073006, [\href{http://arxiv.org/abs/1101.2755}{{\tt 1101.2755}}].

\bibitem{Hayes:2013wra}
A.~C. Hayes, J.~L. Friar, G.~T. Garvey, G.~Jungman, and G.~Jonkmans,
  \href{http://dx.doi.org/10.1103/PhysRevLett.112.202501}{{\it {Systematic
  Uncertainties in the Analysis of the Reactor Neutrino Anomaly}}, } {\em Phys.
  Rev. Lett.} {\bf 112} (2014) 202501,
  [\href{http://arxiv.org/abs/1309.4146}{{\tt 1309.4146}}].

\bibitem{Hayes:2016qnu}
A.~C. Hayes and P.~Vogel,
  \href{http://dx.doi.org/10.1146/annurev-nucl-102115-044826}{{\it {Reactor
  Neutrino Spectra}}, } {\em Ann. Rev. Nucl. Part. Sci.} {\bf 66} (2016)
  219--244, [\href{http://arxiv.org/abs/1605.02047}{{\tt 1605.02047}}].

\bibitem{Vogel:2016ted}
P.~Vogel,
  \href{http://inspirehep.net/record/1436503/files/arXiv:1603.08990.pdf}{{\it
  {Evaluation of Reactor Neutrino Flux: Issues and Uncertainties}}, } in {\em
  {Prospects in Neutrino Physics (Nuphy$S^2$015) London, Uk, December 16-18,
  2015}}, 2016.
\newblock \href{http://arxiv.org/abs/1603.08990}{{\tt 1603.08990}}.

\bibitem{Huber:2016fkt}
P.~Huber, \href{http://dx.doi.org/10.1016/j.nuclphysb.2016.04.012}{{\it
  {Reactor Antineutrino Fluxes -- Status and Challenges}}, } {\em Nucl. Phys.}
  {\bf B908} (2016) 268--278, [\href{http://arxiv.org/abs/1602.01499}{{\tt
  1602.01499}}].

\bibitem{Hayes:2017res}
A.~Hayes, G.~Jungman, L.~McCutchan, A.~Sonzogni, G.~Garvey, et~al., {\it
  {Analysis of the Daya Bay Reactor Antineutrino Flux Changes with Fuel
  Burnup}},  \href{http://arxiv.org/abs/1707.07728}{{\tt 1707.07728}}.

\bibitem{Acero:2007su}
M.~A. Acero, C.~Giunti, and M.~Laveder,
  \href{http://dx.doi.org/10.1103/PhysRevD.78.073009}{{\it {Limits on $\nu_e$
  and $\bar\nu_e$ disappearance from Gallium and reactor experiments}}, } {\em
  Phys.Rev.} {\bf D78} (2008) 073009,
  [\href{http://arxiv.org/abs/0711.4222}{{\tt 0711.4222}}].

\bibitem{Giunti:2010zu}
C.~Giunti and M.~Laveder,
  \href{http://dx.doi.org/10.1103/PhysRevC.83.065504}{{\it {Statistical
  Significance of the Gallium Anomaly}}, } {\em Phys.Rev.} {\bf C83} (2011)
  065504, [\href{http://arxiv.org/abs/1006.3244}{{\tt 1006.3244}}].

\bibitem{Kopp:2011qd}
J.~Kopp, M.~Maltoni, and T.~Schwetz,
  \href{http://dx.doi.org/10.1103/PhysRevLett.107.091801}{{\it {Are there
  sterile neutrinos at the eV scale?}}, } {\em Phys.Rev.Lett.} {\bf 107} (2011)
  091801, [\href{http://arxiv.org/abs/1103.4570}{{\tt 1103.4570}}].

\bibitem{Kopp:2013vaa}
J.~Kopp, P.~A.~N. Machado, M.~Maltoni, and T.~Schwetz,
  \href{http://dx.doi.org/10.1007/JHEP05(2013)050}{{\it {Sterile Neutrino
  Oscillations: the Global Picture}}, } {\em JHEP} {\bf 05} (2013) 050,
  [\href{http://arxiv.org/abs/1303.3011}{{\tt 1303.3011}}].

\bibitem{Gariazzo:2017fdh}
S.~Gariazzo, C.~Giunti, M.~Laveder, and Y.~F. Li,
  \href{http://dx.doi.org/10.1007/JHEP06(2017)135}{{\it {Updated Global 3+1
  Analysis of Short-Baseline Neutrino Oscillations}}, } {\em JHEP} {\bf 06}
  (2017) 135, [\href{http://arxiv.org/abs/1703.00860}{{\tt 1703.00860}}].

\bibitem{Collin:2016aqd}
G.~H. Collin, C.~A. Arg{\"u}elles, J.~M. Conrad, and M.~H. Shaevitz, {\it
  {First Constraints on the Complete Neutrino Mixing Matrix with a Sterile
  Neutrino}},  \href{http://arxiv.org/abs/1607.00011}{{\tt 1607.00011}}.

\bibitem{Seo:2016uom}
H.~Seo et~al., {\it {Spectral Measurement of the Electron Antineutrino
  Oscillation Amplitude and Frequency Using 500 Live Days of Reno Data}},
  \href{http://arxiv.org/abs/1610.04326}{{\tt 1610.04326}}.

\bibitem{Abe:2014bwa}
{\bf Double Chooz}, Y.~Abe et~al.,
  \href{http://dx.doi.org/10.1007/JHEP02(2015)074;
  10.1007/JHEP10(2014)086}{{\it {Improved Measurements of the Neutrino Mixing
  Angle $\theta_{13}$ with the Double Chooz Detector}}, } {\em JHEP} {\bf 10}
  (2014) 086, [\href{http://arxiv.org/abs/1406.7763}{{\tt 1406.7763}}].
  [Erratum: JHEP02,074(2015)].

\bibitem{An:2016srz}
{\bf Daya Bay}, F.~P. An et~al.,
  \href{http://dx.doi.org/10.1088/1674-1137/41/1/013002}{{\it {Improved
  Measurement of the Reactor Antineutrino Flux and Spectrum at Daya Bay}}, }
  {\em Chin. Phys.} {\bf C41} (2017), no.~1 013002,
  [\href{http://arxiv.org/abs/1607.05378}{{\tt 1607.05378}}].

\bibitem{Mention:2017dyq}
G.~Mention, M.~Vivier, J.~Gaffiot, T.~Lasserre, A.~Letourneau, et~al., {\it
  {Reactor Antineutrino Shoulder Explained by Energy Scale Nonlinearities?}},
  \href{http://arxiv.org/abs/1705.09434}{{\tt 1705.09434}}.

\bibitem{Huber:2016xis}
P.~Huber, \href{http://dx.doi.org/10.1103/PhysRevLett.118.042502}{{\it {Neos
  Data and the Origin of the 5 Mev Bump in the Reactor Antineutrino Spectrum}},
  } {\em Phys. Rev. Lett.} {\bf 118} (2017), no.~4 042502,
  [\href{http://arxiv.org/abs/1609.03910}{{\tt 1609.03910}}].

\bibitem{Hayes:2015yka}
A.~C. Hayes, J.~L. Friar, G.~T. Garvey, D.~Ibeling, G.~Jungman, et~al.,
  \href{http://dx.doi.org/10.1103/PhysRevD.92.033015}{{\it {Possible Origins
  and Implications of the Shoulder in Reactor Neutrino Spectra}}, } {\em Phys.
  Rev.} {\bf D92} (2015), no.~3 033015,
  [\href{http://arxiv.org/abs/1506.00583}{{\tt 1506.00583}}].

\bibitem{An:2016luf}
{\bf Daya Bay}, F.~P. An et~al.,
  \href{http://dx.doi.org/10.1103/PhysRevLett.117.151802}{{\it {Improved Search
  for a Light Sterile Neutrino with the Full Configuration of the Daya Bay
  Experiment}}, } {\em Phys. Rev. Lett.} {\bf 117} (2016), no.~15 151802,
  [\href{http://arxiv.org/abs/1607.01174}{{\tt 1607.01174}}].

\bibitem{Ko:2016owz}
Y.~J. Ko et~al., \href{http://dx.doi.org/10.1103/PhysRevLett.118.121802}{{\it
  {Sterile Neutrino Search at the Neos Experiment}}, } {\em Phys. Rev. Lett.}
  {\bf 118} (2017), no.~12 121802, [\href{http://arxiv.org/abs/1610.05134}{{\tt
  1610.05134}}].

\bibitem{Alekseev:2016llm}
I.~Alekseev et~al.,
  \href{http://dx.doi.org/10.1088/1748-0221/11/11/P11011}{{\it {Danss: Detector
  of the Reactor Antineutrino Based on Solid Scintillator}}, } {\em JINST} {\bf
  11} (2016), no.~11 P11011, [\href{http://arxiv.org/abs/1606.02896}{{\tt
  1606.02896}}].

\bibitem{danss-moriond17}
M.~Danilov, {\it {Search for sterile neutrinos at the DANSS and Neutrino-4
  experiments}},  2017.
\newblock talk given on behalf of the DANSS Collaboration at the 52nd
  Rencontres de Moriond EW 2017, La Thuile, Italy; {\tt
  https://indico.in2p3.fr/event/13763/}.

\bibitem{An:2017osx}
{\bf Daya Bay}, F.~P. An et~al.,
  \href{http://dx.doi.org/10.1103/PhysRevLett.118.251801}{{\it {Evolution of
  the Reactor Antineutrino Flux and Spectrum at Daya Bay}}, } {\em Phys. Rev.
  Lett.} {\bf 118} (2017), no.~25 251801,
  [\href{http://arxiv.org/abs/1704.01082}{{\tt 1704.01082}}].

\bibitem{Giunti:2016elf}
C.~Giunti, \href{http://dx.doi.org/10.1016/j.physletb.2016.11.028}{{\it
  {Precise Determination of the $^{235}$U Reactor Antineutrino Cross Section
  Per Fission}}, } {\em Phys. Lett.} {\bf B764} (2017) 145--149,
  [\href{http://arxiv.org/abs/1608.04096}{{\tt 1608.04096}}].

\bibitem{Giunti:2017nww}
C.~Giunti, {\it {Improved Determination of the $^{235}\text{U}$ and
  $^{239}\text{Pu}$ Reactor Antineutrino Cross Sections per Fission}},
  \href{http://arxiv.org/abs/1704.02276}{{\tt 1704.02276}}.

\bibitem{Giunti:2017yid}
C.~Giunti, X.~P. Ji, M.~Laveder, Y.~F. Li, and B.~R. Littlejohn, {\it {Reactor
  Fuel Fraction Information on the Antineutrino Anomaly}},
  \href{http://arxiv.org/abs/1708.01133}{{\tt 1708.01133}}.

\bibitem{Aguilar:2001ty}
{\bf LSND}, A.~Aguilar et~al., {\it {Evidence for neutrino oscillations from
  the observation of $\bar{\nu}_e$ appearance in a $\bar{\nu}_\mu$ beam}},
  {\em Phys. Rev.} {\bf D64} (2001) 112007,
  [\href{http://arxiv.org/abs/hep-ex/0104049}{{\tt hep-ex/0104049}}].

\bibitem{Kozlov:1999cs}
{\relax Yu}.~V. Kozlov et~al.,
  \href{http://dx.doi.org/10.1016/S0920-5632(00)00738-6}{{\it {Today and Future
  Neutrino Experiments at Krasnoyarsk Nuclear Reactor}}, } {\em Nucl. Phys.
  Proc. Suppl.} {\bf 87} (2000) 514--516,
  [\href{http://arxiv.org/abs/hep-ex/9912046}{{\tt hep-ex/9912046}}].

\bibitem{reno-EPS17}
H.~Seo, {\it {New results from RENO}},  2017.
\newblock talk given on behalf of the RENO Collaboration at the EPS conference
  on High Energy Physics, Venice, Italy, July 5--11, 2017.

\bibitem{reno-Neutrino14}
S.-H. Seo, {\it {Results from RENO}},  2014.
\newblock talk given on behalf of the RENO Collaboration at the XXVI
  International Conference on Neutrino Physics and Astrophysics, Boston, USA,
  June 2--7, 2014.

\bibitem{Declais:1994ma}
Y.~Declais et~al., \href{http://dx.doi.org/10.1016/0370-2693(94)91394-3}{{\it
  {Study of reactor anti-neutrino interaction with proton at Bugey nuclear
  power plant}}, } {\em Phys.Lett.} {\bf B338} (1994) 383--389.

\bibitem{Kwon:1981ua}
H.~Kwon et~al., \href{http://dx.doi.org/10.1103/PhysRevD.24.1097}{{\it {Search
  for neutrino oscillations at a fission reactor}}, } {\em Phys.Rev.} {\bf D24}
  (1981) 1097--1111.

\bibitem{Zacek:1986cu}
G.~Zacek et~al., \href{http://dx.doi.org/10.1103/PhysRevD.34.2621}{{\it
  {Neutrino Oscillation Experiments at the Gosgen Nuclear Power Reactor}}, }
  {\em Phys.Rev.} {\bf D34} (1986) 2621--2636.

\bibitem{Vidyakin:1987ue}
G.~Vidyakin et~al., {\it {Detection of anti-neutrinos in the flux from two
  reactors}},  {\em Sov.Phys.JETP} {\bf 66} (1987) 243--247.

\bibitem{Vidyakin:1994ut}
G.~S. Vidyakin et~al., {\it {Limitations on the characteristics of neutrino
  oscillations}},  {\em JETP Lett.} {\bf 59} (1994) 390--393.

\bibitem{Afonin:1988gx}
A.~Afonin, S.~Ketov, V.~Kopeikin, L.~Mikaelyan, M.~Skorokhvatov, et~al., {\it
  {A study of the reaction $\bar\nu_e + p \to e^+ + n$ on a nuclear reactor}},
  {\em Sov.Phys.JETP} {\bf 67} (1988) 213--221.

\bibitem{Kuvshinnikov:1990ry}
A.~Kuvshinnikov et~al., {\it {Measuring the anti-electron-neutrino + p $\to$ n
  + e+ cross-section and beta decay axial constant in a new experiment at Rovno
  NPP reactor.}},  {\em JETP Lett.} {\bf 54} (1991) 253--257.

\bibitem{Greenwood:1996pb}
Z.~D. Greenwood et~al., \href{http://dx.doi.org/10.1103/PhysRevD.53.6054}{{\it
  {Results of a two position reactor neutrino oscillation experiment}}, } {\em
  Phys. Rev.} {\bf D53} (1996) 6054--6064.

\bibitem{Declais:1994su}
Y.~Declais et~al., \href{http://dx.doi.org/10.1016/0550-3213(94)00513-E}{{\it
  {Search for neutrino oscillations at 15-meters, 40-meters, and 95-meters from
  a nuclear power reactor at Bugey}}, } {\em Nucl.Phys.} {\bf B434} (1995)
  503--534.

\bibitem{An:2016ses}
{\bf Daya Bay}, F.~P. An et~al.,
  \href{http://dx.doi.org/10.1103/PhysRevD.95.072006}{{\it {Measurement of
  Electron Antineutrino Oscillation Based on 1230 Days of Operation of the Daya
  Bay Experiment}}, } {\em Phys. Rev.} {\bf D95} (2017), no.~7 072006,
  [\href{http://arxiv.org/abs/1610.04802}{{\tt 1610.04802}}].

\bibitem{Gando:2010aa}
{\bf KamLAND}, A.~Gando et~al.,
  \href{http://dx.doi.org/10.1103/PhysRevD.83.052002}{{\it {Constraints on
  $\theta_{13}$ from A Three-Flavor Oscillation Analysis of Reactor
  Antineutrinos at KamLAND}}, } {\em Phys.Rev.} {\bf D83} (2011) 052002,
  [\href{http://arxiv.org/abs/1009.4771}{{\tt 1009.4771}}].

\bibitem{Grimus:2001mn}
W.~Grimus and T.~Schwetz, \href{http://dx.doi.org/10.1007/s100520100646}{{\it
  {4-neutrino mass schemes and the likelihood of (3+1)-mass spectra}}, } {\em
  Eur. Phys. J.} {\bf C20} (2001) 1--11,
  [\href{http://arxiv.org/abs/hep-ph/0102252}{{\tt hep-ph/0102252}}].

\bibitem{Cleveland:1998nv}
B.~T. Cleveland et~al., \href{http://dx.doi.org/10.1086/305343}{{\it
  {Measurement of the solar electron neutrino flux with the Homestake chlorine
  detector}}, } {\em Astrophys. J.} {\bf 496} (1998) 505--526.

\bibitem{Kaether:2010ag}
F.~Kaether, W.~Hampel, G.~Heusser, J.~Kiko, and T.~Kirsten,
  \href{http://dx.doi.org/10.1016/j.physletb.2010.01.030}{{\it {Reanalysis of
  the GALLEX solar neutrino flux and source experiments}}, } {\em Phys.Lett.}
  {\bf B685} (2010) 47--54, [\href{http://arxiv.org/abs/1001.2731}{{\tt
  1001.2731}}].

\bibitem{Abdurashitov:2009tn}
{\bf SAGE}, J.~N. Abdurashitov et~al.,
  \href{http://dx.doi.org/10.1103/PhysRevC.80.015807}{{\it {Measurement of the
  solar neutrino capture rate with gallium metal. III: Results for the
  2002--2007 data-taking period}}, } {\em Phys. Rev.} {\bf C80} (2009) 015807,
  [\href{http://arxiv.org/abs/0901.2200}{{\tt 0901.2200}}].

\bibitem{Hosaka:2005um}
{\bf Super-Kamiokande}, J.~Hosaka et~al.,
  \href{http://dx.doi.org/10.1103/PhysRevD.73.112001}{{\it {Solar neutrino
  measurements in Super-Kamiokande-I}}, } {\em Phys. Rev.} {\bf D73} (2006)
  112001, [\href{http://arxiv.org/abs/hep-ex/0508053}{{\tt hep-ex/0508053}}].

\bibitem{Cravens:2008aa}
{\bf Super-Kamiokande}, J.~Cravens et~al.,
  \href{http://dx.doi.org/10.1103/PhysRevD.78.032002}{{\it {Solar neutrino
  measurements in Super-Kamiokande-II}}, } {\em Phys.Rev.} {\bf D78} (2008)
  032002, [\href{http://arxiv.org/abs/0803.4312}{{\tt 0803.4312}}].

\bibitem{Abe:2010hy}
{\bf Super-Kamiokande}, K.~Abe et~al.,
  \href{http://dx.doi.org/10.1103/PhysRevD.83.052010}{{\it {Solar neutrino
  results in Super-Kamiokande-III}}, } {\em Phys.Rev.} {\bf D83} (2011) 052010,
  [\href{http://arxiv.org/abs/1010.0118}{{\tt 1010.0118}}].

\bibitem{sksol:nakano2016}
Y.~Nakano, {\em {$^8$B solar neutrino spectrum measurement using
  Super-Kamiokande IV}}.
\newblock PhD thesis, Tokyo U., 2016-02.

\bibitem{Aharmim:2007nv}
{\bf SNO}, B.~Aharmim et~al.,
  \href{http://dx.doi.org/10.1103/PhysRevC.75.045502}{{\it {Measurement of the
  $\nu_e$ and total B-8 solar neutrino fluxes with the Sudbury Neutrino
  Observatory phase I data set}}, } {\em Phys. Rev.} {\bf C75} (2007) 045502,
  [\href{http://arxiv.org/abs/nucl-ex/0610020}{{\tt nucl-ex/0610020}}].

\bibitem{Aharmim:2005gt}
{\bf SNO}, B.~Aharmim et~al.,
  \href{http://dx.doi.org/10.1103/PhysRevC.72.055502}{{\it {Electron energy
  spectra, fluxes, and day-night asymmetries of B-8 solar neutrinos from the
  391-day salt phase SNO data set}}, } {\em Phys. Rev.} {\bf C72} (2005)
  055502, [\href{http://arxiv.org/abs/nucl-ex/0502021}{{\tt nucl-ex/0502021}}].

\bibitem{Aharmim:2008kc}
{\bf SNO}, B.~Aharmim et~al.,
  \href{http://dx.doi.org/10.1103/PhysRevLett.101.111301}{{\it {An Independent
  Measurement of the Total Active 8B Solar Neutrino Flux Using an Array of 3He
  Proportional Counters at the Sudbury Neutrino Observatory}}, } {\em Phys.
  Rev. Lett.} {\bf 101} (2008) 111301,
  [\href{http://arxiv.org/abs/0806.0989}{{\tt 0806.0989}}].

\bibitem{Bellini:2011rx}
{\bf Borexino}, G.~Bellini et~al.,
  \href{http://dx.doi.org/10.1103/PhysRevLett.107.141302}{{\it {Precision
  measurement of the 7Be solar neutrino interaction rate in Borexino}}, } {\em
  Phys.Rev.Lett.} {\bf 107} (2011) 141302,
  [\href{http://arxiv.org/abs/1104.1816}{{\tt 1104.1816}}].

\bibitem{Bellini:2008mr}
{\bf Borexino}, G.~Bellini et~al.,
  \href{http://dx.doi.org/10.1103/PhysRevD.82.033006}{{\it {Measurement of the
  solar 8B neutrino rate with a liquid scintillator target and 3 MeV energy
  threshold in the Borexino detector}}, } {\em Phys.Rev.} {\bf D82} (2010)
  033006, [\href{http://arxiv.org/abs/0808.2868}{{\tt 0808.2868}}].

\bibitem{Bellini:2014uqa}
{\bf Borexino}, G.~Bellini et~al.,
  \href{http://dx.doi.org/10.1038/nature13702}{{\it {Neutrinos from the primary
  proton--proton fusion process in the Sun}}, } {\em Nature} {\bf 512} (2014),
  no.~7515 383--386.

\bibitem{Hampel:1997fc}
{\bf GALLEX}, W.~Hampel et~al.,
  \href{http://dx.doi.org/10.1016/S0370-2693(97)01562-1}{{\it {Final results of
  the Cr-51 neutrino source experiments in GALLEX}}, } {\em Phys.Lett.} {\bf
  B420} (1998) 114--126.

\bibitem{Abdurashitov:1998ne}
{\bf SAGE}, J.~Abdurashitov et~al.,
  \href{http://dx.doi.org/10.1103/PhysRevC.59.2246}{{\it {Measurement of the
  response of the Russian-American gallium experiment to neutrinos from a Cr-51
  source}}, } {\em Phys.Rev.} {\bf C59} (1999) 2246--2263,
  [\href{http://arxiv.org/abs/hep-ph/9803418}{{\tt hep-ph/9803418}}].

\bibitem{Abdurashitov:2005tb}
J.~Abdurashitov, V.~Gavrin, S.~Girin, V.~Gorbachev, P.~Gurkina, et~al.,
  \href{http://dx.doi.org/10.1103/PhysRevC.73.045805}{{\it {Measurement of the
  response of a Ga solar neutrino experiment to neutrinos from an Ar-37
  source}}, } {\em Phys.Rev.} {\bf C73} (2006) 045805,
  [\href{http://arxiv.org/abs/nucl-ex/0512041}{{\tt nucl-ex/0512041}}].

\bibitem{Reichenbacher:2005nc}
J.~Reichenbacher, {\it {Final KARMEN results on neutrino oscillations and
  neutrino nucleus interactions in the energy regime of supernovae}}, . {PhD
  thesis, Univ.\ Karlsruhe}.

\bibitem{Armbruster:1998uk}
B.~Armbruster, I.~Blair, B.~Bodmann, N.~Booth, G.~Drexlin, et~al.,
  \href{http://dx.doi.org/10.1103/PhysRevC.57.3414}{{\it {KARMEN limits on
  electron-neutrino $\to$ tau-neutrino oscillations in two neutrino and three
  neutrino mixing schemes}}, } {\em Phys.Rev.} {\bf C57} (1998) 3414--3424,
  [\href{http://arxiv.org/abs/hep-ex/9801007}{{\tt hep-ex/9801007}}].

\bibitem{Conrad:2011ce}
J.~Conrad and M.~Shaevitz,
  \href{http://dx.doi.org/10.1103/PhysRevD.85.013017}{{\it {Limits on Electron
  Neutrino Disappearance from the KARMEN and LSND $\nu_e$ - Carbon Cross
  Section Data}}, } {\em Phys.Rev.} {\bf D85} (2012) 013017,
  [\href{http://arxiv.org/abs/1106.5552}{{\tt 1106.5552}}].

\bibitem{Auerbach:2001hz}
{\bf LSND}, L.~Auerbach et~al.,
  \href{http://dx.doi.org/10.1103/PhysRevC.64.065501}{{\it {Measurements of
  charged current reactions of nu(e) on 12-C}}, } {\em Phys.Rev.} {\bf C64}
  (2001) 065501, [\href{http://arxiv.org/abs/hep-ex/0105068}{{\tt
  hep-ex/0105068}}].

\bibitem{Frekers:2011zz}
D.~Frekers, H.~Ejiri, H.~Akimune, T.~Adachi, B.~Bilgier, et~al., {\it {The
  Ga-71(He-3, t) reaction and the low-energy neutrino response}},  {\em
  Phys.Lett.} {\bf B706} (2011) 134--138.

\bibitem{Bahcall:1997eg}
J.~N. Bahcall, \href{http://dx.doi.org/10.1103/PhysRevC.56.3391}{{\it {Gallium
  solar neutrino experiments: Absorption cross-sections, neutrino spectra, and
  predicted event rates}}, } {\em Phys.Rev.} {\bf C56} (1997) 3391--3409,
  [\href{http://arxiv.org/abs/hep-ph/9710491}{{\tt hep-ph/9710491}}].

\bibitem{Vogel:1999zy}
P.~Vogel and J.~F. Beacom, {\it {The angular distribution of the neutron
  inverse beta decay, $\bar{\nu}_e + p \rightarrow e^+ + n$}},  {\em Phys.
  Rev.} {\bf D60} (1999) 053003,
  [\href{http://arxiv.org/abs/hep-ph/9903554}{{\tt hep-ph/9903554}}].

\bibitem{Danilov:2014vra}
{\bf DANSS}, M.~Danilov, {\it {Sensitivity of DANSS detector to short range
  neutrino oscillations}},  \href{http://arxiv.org/abs/1412.0817}{{\tt
  1412.0817}}.

\end{thebibliography}\endgroup

\end{document}